\documentclass[prd,twocolumn,floatfix,nofootinbib]{revtex4}
\usepackage{amssymb}
\usepackage{amsmath}
\usepackage{graphicx,subfigure,color,dcolumn,booktabs,bm}
\usepackage{longtable,lscape}
\usepackage{txfonts}
\usepackage{overpic}
\usepackage{indentfirst}
\usepackage{cases}
\usepackage{multirow}
\usepackage{ulem}
\usepackage{enumerate}
\usepackage[colorlinks,
            citecolor=blue,
            anchorcolor=red,
            menucolor=red,
            linkcolor=red,
            filecolor=red,
            runcolor=red,
            urlcolor=blue,
            frenchlinks=false]{hyperref}

\allowdisplaybreaks
\begin{document}

\title{Termination of the Radial Ladder in Mixed-Flavor Heavy Mesons}
\author{Wen-Xuan Zhang$^{1,2,3,4}$}
\email{zhangwx89@outlook.com}
\author{Wen-Nian Liu$^{5}$}
\email{phyliu1992@yeah.net}
\affiliation{$^1$School of Physical Science and Technology, Lanzhou University, Lanzhou 730000, China \\
$^2$Lanzhou Center for Theoretical Physics, Key Laboratory of Theoretical Physics of Gansu Province,
Key Laboratory of Quantum Theory and Applications of MoE,
Gansu Provincial Research Center for Basic Disciplines of Quantum Physics, Lanzhou University, Lanzhou 730000, China\\
$^3$MoE Frontiers Science Center for Rare Isotopes, Lanzhou University, Lanzhou 730000, China\\
$^4$Research Center for Hadron and CSR Physics, Lanzhou University and Institute of Modern Physics of CAS, Lanzhou 730000, China\\
$^5$Institute of Theoretical Physics, College of Physics and Electronic Engineering, Northwest Normal University, Lanzhou 730070, China}

\begin{abstract}
A fundamental open question in hadron spectroscopy is whether the radial excitation ladder of heavy mesons terminates at a finite level once confinement is screened by light-quark pair creation. We address this question for the mixed-flavor mesons $D_s$, $B_s$, and $B_c$ by solving a screened Godfrey--Isgur Hamiltonian with the Gaussian expansion method. The screening length is fixed by threshold-dependent string breaking analysis, and is matched to the charmonium and bottomonium anchors. The calculated spectra saturate at $3.36$, $6.78$, and $8.19~{\rm GeV}$ in $D_s$, $B_s$, and $B_c$ families, while the RMS radii grow by several fm and mass gaps fall below $10~{\rm MeV}$. Mass-radius decoupling places the end of the safe radial ladder at $n_{\mathrm{safe}}\approx 4$--$6$ for $D_s$, $5$--$7$ for $B_s$, and $9$--$11$ for $B_c$, so the ladder terminates first in $D_s$, then in $B_s$, and last in $B_c$. Above the limiting-mass windows, at $3.49$, $6.86$, and $8.24~{\rm GeV}$, states of the radial ladder cannot be discovered. This termination is directly testable with forthcoming high-statistics data from BESIII, Belle II, and LHCb.
\end{abstract}

\maketitle

\section{Introduction}
\label{sec:introduction}

A quantitative description of color confinement, central to the nonperturbative strong interaction, remains an important issue in particle physics. Hadron spectroscopy, approached through phenomenology, experiment, and lattice QCD, provides a direct probe of this nonperturbative dynamics. The Cornell potential, $V(r)=ar+b/r$, successfully describes the low-lying meson spectrum with linear confinment in a quenched picture of the flux tube~\cite{Eichten:1974af,Eichten:1978tg,Eichten:1979ms}. This potential continues to rise at long distance, implies an unbounded mass spectrum and the radial ladder built on it has no upper limit. The unquenched effects resolve this limitation and modify the long-distance behavior of confinement by creating a light quark-antiquark pair and breaking the color string~\cite{Bali:2005fu,Castorina:2007eb,Bulava:2019iut,Jiang:2023lmj,Kou:2024dml}. The radial ladder of a heavy meson is then expected to terminate at a finite mass. The present work raises up this question for the mixed-flavor mesons $D_s$, $B_s$, and $B_c$.

Experimental study of $D_s$, $B_s$, and $B_c$ now reaches the first radial and orbital excitations~\cite{Chen:2017rpp,ParticleDataGroup:2024cfk}. BaBar observed the narrow $D_{s0}^{*}(2317)$ in $D_s\pi^{0}$ below the $DK$ threshold \cite{BaBar:2003ds2317}. CLEO and Belle confirmed that state and established $D_{s1}(2460)$ below the $D^{*}K$ threshold~\cite{CLEO:2003ds2460,Belle:2003ds2457}. The narrow $P$-wave states $D_{s1}(2536)$ and $D_{s2}^{*}(2573)$ have already been observed~\cite{ARGUS:1989ds1,CLEO:1994ds2}. Belle then reported a structure $D_{s1}^{*}(2700)$ in $DK$~\cite{Belle:2008ds2700}. LHCb resolved $D_{s1}^{*}(2860)$ and $D_{s3}^{*}(2860)$ as overlapping spin-1 and spin-3 resonances~\cite{LHCb:2014ds2860}. It later observed $D_{s0}(2590)$ and $D_{s1}(2933)$~\cite{LHCb:2020gey,LHCb:2026ds2933}. In the $B_s$ family, CDF observed the narrow states $B_{s1}(5830)$ and $B_{s2}^{*}(5840)$, and D0 observed $B_{s2}^{*}(5840)$~\cite{CDF:2008bs,D0:2008bs}. LHCb later reported $B_{sJ}(6063)$ and $B_{sJ}(6114)$~\cite{LHCb:2020bsJ}. In the $B_c$ family, ATLAS reported an excited peak that CMS resolved into $B_c(2S)$ and $B_c^{*}(2S)$~\cite{ATLAS:2014bc2S,CMS:2019bc2S}. ATLAS observed the hyperfine partner $B_c^{*}$ of the ground state~\cite{ATLAS:2026abc}. LHCb reported the orbital candidates $B_c(6700)$ and $B_c(6750)$~\cite{LHCb:2025bcP}. Each family is therefore known at the bottom of its radial ladder, while the discoveries from higher energy sector remain absent.

The same families have been studied in relativized quark models, from the Godfrey--Isgur Hamiltonian through later calculations of the $D_s$, $B_s$, and $B_c$ excitation spectra~\cite{Godfrey:1985xj,DiPierro:2001hl,Godfrey:2004bc,Godfrey:2016charm,Godfrey:2016bs,Ebert:2002pp,Ebert:2009ua,Ni:2022cs}. With a linear confining term, these models describe the established ground states and the narrow $P$-wave levels. They place the assignments of $D_{s0}^{*}(2317)$ and $D_{s1}(2460)$ above the masses observed below the $DK$ and $D^{*}K$ thresholds~\cite{Ebert:2009ua,Ni:2022cs}. To solve this mismatch, coupled-channel analysis of the scalar $c\bar s$ state to the $DK$ continuum shifts the mass downwards~\cite{Luo:2021dvj,Zhang:2024usz}, accounting for spectrum dressing from the unquenched effects. Screened confinement encodes the same unquenching by saturating the linear rise, and it has been applied to heavy-meson masses and strong decays~\cite{Gonzalez:2003scr,Song:2015nia,Wang:2018rjg,Wang:2019mhs,Wang:2020prx}. The unquenched effects are essential for describing the high-lying spectrum~\cite{Bai:2026atm}. They lead to meson energy levels that differ from those obtained in the quenched approximation~\cite{Ding:1993uy,Li:2009nr,Deng:2023mza,Chen:2024ukv}. This implies that the meson spectrum is strongly constrained at large radial number.

This work is directly inspired from previous work on the termination of the radial ladder in heavy quarkonia, mainly for charmonium~\cite{Zhang:2026yif}. We determine that termination here for the mixed-flavor systems $D_s$ ($c\bar s$), $B_s$ ($b\bar s$), and $B_c$ ($b\bar c$). The spectrum of each family is obtained from a screen-modified Godfrey--Isgur (GI) Hamiltonian \cite{Godfrey:1985xj,Capstick:1986ter}, solved with the Gaussian expansion method (GEM)~\cite{Hiyama:2003cu,Hiyama:2012sma}. The screening parameter is fixed by threshold-dependent string breaking through light-quark pair creation, and is matched to the charmonium and bottomonium anchors. With that parameter held fixed, the string tension and the relativistic exponents are determined from the established states that carry a unique spectroscopic assignment. Masses and root-mean-square (RMS) radii are then computed along the $S$, $P$, and $D$ waves. The end of each ladder is identified with the onset of mass-radius decoupling, where the mass approaches a plateau while the RMS radius continues to grow. The calculated spectra saturate at a finite mass in each family. The corresponding safe radial range is shortest in $D_s$ and longest in $B_c$, so the ladder terminates first in $D_s$, then in $B_s$, and last in $B_c$. Above those finite-mass windows, further states of the radial ladder cannot be discovered. 

This work is structured as follows. In Sec.~\ref{sec:framework}, we specify the screened Godfrey--Isgur Hamiltonian and the Gaussian expansion method used to solve it. Sec.~\ref{sec:screening} determines the screening parameter of $D_s$, $B_s$, and $B_c$ from threshold-dependent string breaking, matched to the charmonium and bottomonium anchors. Sec.~\ref{sec:results} presents the fitted parameters and the spectra, and then locates the termination of each radial ladder. A summary of conclusions is given in Sec.~\ref{sec:summary}. The extended mass and radius tables, together with the termination diagnostics, are collected in Appendix~\ref{app:spectra}.

\section{Theoretical framework}
\label{sec:framework}

The spectroscopic analysis is built on a screen-modified GI Hamiltonian, solved with the GEM that can represent both low-lying states and highly extended radial excitations. The systems of interest are the mixed-flavor heavy mesons $D_s$ ($c\bar s$), $B_s$ ($b\bar s$), and $B_c$ ($b\bar c$). We first specify the Hamiltonian, then the Gaussian basis and the evaluation of matrix elements, and finally the RMS radius.

The Hamiltonian takes the form
\begin{equation}
  \begin{aligned}
  H &= \sum_{i}\sqrt{p_i^2+m_i^2} \\
  &+ \sum_{i<j}\Bigl(V_{ij}^{\rm Coul}
  + V_{ij}^{\rm conf}
  + V_{ij}^{\rm cont}
  + V_{ij}^{\rm so(\nu)}
  + V_{ij}^{\rm so(s)}
  + V_{ij}^{\rm tens}\Bigr).
  \end{aligned}
  \label{equ:GIHamiltonium}
\end{equation}
The first sum is the relativistic kinetic energy of each constituent. The following terms are, respectively, the color--Coulomb, screened confinement, contact, spin--orbit, Thomas-precession, and tensor interactions. Among these, the Coulomb and confinement interactions dominate the short-distance binding and the long-range confining mechanism, respectively. The GI construction regularizes the Coulomb singularity by Gaussian smearing and applies the same finite-size smearing to the screened confinement. The other potentials are therefore modified with the smeared Coulomb or confinement kernels.

The color-Coulomb interaction of one-gluon exchange is singular at the origin as $r^{-1}$. Following the GI framework, it is smeared by a Gaussian convolution kernel, which produces the smeared Coulomb potential
\begin{align}
    \widetilde{G}(r) &= \int d^3\mathbf{r^{\prime}} \big(\frac{\sigma^{3}}{\pi^{3/2}}
    e^{-\sigma^2 (\mathbf{r}-\mathbf{r^{\prime}})^2}\big) \big(c\frac{\alpha}{r^{\prime}}\big) \nonumber \\
    &= \iint d\theta dr^{\prime} 2\pi{r^{\prime}}^2\textrm{sin}(\theta) \big(\frac{\sigma^{3}}{\pi^{3/2}}
    e^{-\sigma^2 (r^2+{r^{\prime}}^2-2rr^{\prime}\textrm{cos}(\theta))}\big) \big(c\frac{\alpha}{r^{\prime}}\big) \nonumber \\
    &= c\frac{\alpha}{r} \textrm{erf}(\sigma r). \label{equ:smearingGr}
\end{align}
The explicit form in the GI model reads
\begin{equation}
\widetilde{G}_{ij}
  = C_{ij}\sum_k\frac{\alpha_k}{r_{ij}}\,\mathrm{erf}(\sigma_{kij}r_{ij}),
\label{equ:Gij}
\end{equation}
with color factor $C_{ij}=-4/3$ for mesons and $\sigma_{kij}^{-2}=\sigma_{ij}^{-2}+\gamma_k^{-2}$. The parameters $\alpha_k$ $(0.25,0.15,0.20)$ and $\gamma_k$ $(\frac{1}{2},\frac{1}{2}\sqrt{10},5\sqrt{10})$ reproduce the running coupling $\alpha_s(Q^2)=\sum_k\alpha_k e^{-Q^2/4\gamma_k^2}$ in Refs.\cite{Godfrey:1985xj,Capstick:1986ter}. The smearing width $\sigma_{ij}$ is not a universal constant, but rather depends on the parameters $\sigma_0$ and $s$ and on the two constituent masses:
\begin{equation}
\sigma_{ij}^2
  = \sigma_0^2\Bigl[\tfrac12+\tfrac12\Bigl(\frac{4m_im_j}{(m_i+m_j)^2}\Bigr)^4\Bigr]
    +s^2\Bigl(\frac{2m_im_j}{m_i+m_j}\Bigr)^2,
\label{equ:sigmaij}
\end{equation}
so that the short-distance Coulomb interaction is $ij$--pair dependent, and is no longer a universal function of $r$.

Smearing alone still evaluates the interaction as if the quarks were at rest. Finite-momentum and relativistic corrections are included through the factors
\begin{align}
\beta_{ij}
  &= 1+\frac{p_{ij}^2}{(p_{ij}^2+m_i^2)^{1/2}(p_{ij}^2+m_j^2)^{1/2}},
\label{equ:beta}\\
\delta_{ij}
  &= \frac{m_im_j}{(p_{ij}^2+m_i^2)^{1/2}(p_{ij}^2+m_j^2)^{1/2}},
\label{equ:delta}\\
\delta_{ii}
  &= \frac{m_im_i}{(p_{ii}^2+m_i^2)^{1/2}(p_{ii}^2+m_i^2)^{1/2}},
\label{equ:deltaii}\\
\delta_{jj}
  &= \frac{m_jm_j}{(p_{jj}^2+m_j^2)^{1/2}(p_{jj}^2+m_j^2)^{1/2}},
\label{equ:deltajj}
\end{align}
where $p_{ij}$ is the momentum of either of the quarks in the $ij$ center-of-mass frame. In the nonrelativistic limit these factors tend to unity, recovering a static potential; at finite momentum they supply the relativistic kinematics that define the GI construction.

With smearing and the relativistic factors in place, the Coulomb term that enters Eq.~(\ref{equ:GIHamiltonium}) is the smeared Coulomb kernel (\ref{equ:Gij}) multiplied on both sides by $\beta_{ij}^{1/2+\epsilon+\epsilon^2}$:
\begin{equation}
V_{ij}^{\rm Coul}
  = \beta_{ij}^{1/2+\epsilon_{\rm Coul}+\epsilon_{\rm Coul}^2}
    \widetilde{G}_{ij}
    \beta_{ij}^{1/2+\epsilon_{\rm Coul}+\epsilon_{\rm Coul}^2}.
\label{equ:VCoul}
\end{equation}
The remaining one-gluon-exchange interactions are generated from the same Coulomb kernel: its Laplacian supplies the contact interaction, while its radial derivatives provide the spin--orbit and tensor interactions:
\begin{align}
V_{ij}^{\rm cont}
  &= \delta_{ij}^{1/2+\epsilon_{\rm cont}+\epsilon_{\rm cont}^2}
     \frac{2\mathbf{S}_i\cdot\mathbf{S}_j}{3m_im_j}\nabla^2\widetilde{G}_{ij}
     \delta_{ij}^{1/2+\epsilon_{\rm cont}+\epsilon_{\rm cont}^2} \nonumber\\
  &= \delta_{ij}^{1/2+\epsilon_{\rm cont}+\epsilon_{\rm cont}^2}
     \frac{32e^{-\sigma_{kij}^2 r_{ij}^2}\alpha_k\sigma_{kij}^3}{9\sqrt{\pi}m_im_j}
     \mathbf{S}_i\cdot\mathbf{S}_j
     \delta_{ij}^{1/2+\epsilon_{\rm cont}+\epsilon_{\rm cont}^2}, \label{equ:Vcont}\\
V_{ij}^{\rm so(\nu)}
  &= \frac{1}{r_{ij}}\frac{d\widetilde{G}_{ij}}{dr_{ij}}
     \Bigl(\delta_{ii}^{1/2+\epsilon_{\rm so(\nu)}+\epsilon_{\rm so(\nu)}^2}\frac{\mathbf{L}\cdot\mathbf{S}_i}{2m_i^2}
     \delta_{ii}^{1/2+\epsilon_{\rm so(\nu)}+\epsilon_{\rm so(\nu)}^2} \nonumber\\
     &\qquad
     +\delta_{jj}^{1/2+\epsilon_{\rm so(\nu)}+\epsilon_{\rm so(\nu)}^2}\frac{\mathbf{L}\cdot\mathbf{S}_j}{2m_j^2}
     \delta_{jj}^{1/2+\epsilon_{\rm so(\nu)}+\epsilon_{\rm so(\nu)}^2} \nonumber\\
     &\qquad
     +\delta_{ij}^{1/2+\epsilon_{\rm so(\nu)}+\epsilon_{\rm so(\nu)}^2}\frac{\mathbf{L}\cdot\mathbf{S}_i+\mathbf{L}\cdot\mathbf{S}_j}{2m_im_j}
     \delta_{ij}^{1/2+\epsilon_{\rm so(\nu)}+\epsilon_{\rm so(\nu)}^2}\Bigr), \label{equ:Vsonu}\\
V_{ij}^{\rm tens}
  &= \delta_{ij}^{1/2+\epsilon_{\rm tens}+\epsilon_{\rm tens}^2}\frac{1}{3m_im_j}
     \Bigl(\frac{1}{r_{ij}}\frac{d\widetilde{G}_{ij}}{dr_{ij}}-\frac{d^2\widetilde{G}_{ij}}{dr_{ij}^2}\Bigr) \nonumber\\
     &\qquad
     \times \Bigl(\frac{3(\mathbf{S}_i\cdot\mathbf{r}_{ij})(\mathbf{S}_j\cdot\mathbf{r}_{ij})}{r_{ij}^2}
     -\mathbf{S}_i\cdot\mathbf{S}_j\Bigr)
     \delta_{ij}^{1/2+\epsilon_{\rm tens}+\epsilon_{\rm tens}^2}. \label{equ:Vtens}
\end{align}
The contact term generates the $S$-wave hyperfine splittings, while the spin--orbit and tensor terms control the remaining fine structure. In Eqs.~(\ref{equ:VCoul})--(\ref{equ:Vtens}) each relativistic factor is raised to the power of $\tfrac12+\epsilon+\epsilon^2$. The leading $\tfrac12$ is the GI default, $\epsilon$ is a phenomenological parameter that controls the strength of the relativistic modification, and the quadratic term is motivated by a Taylor expansion in $\epsilon$.

The Coulomb, contact, spin--orbit, and tensor interactions all follow from one-gluon exchange. They control the short-distance binding and the fine structure, but they fall off at large $r$ and cannot terminate the radial ladder. Only a confining potential that is screened at large $r$ can do so. In place of the linear confining potential of the original GI model, we use the screened form~\cite{Ding:1993uy,Song:2015nia,Wang:2018rjg,Wang:2019mhs}
\begin{equation}
  S(r)=\frac{b}{\mu}\bigl(1-e^{-\mu r}\bigr)+c.
  \label{equ:Vconf}
\end{equation}
As $r\to\infty$ this potential saturates at $b/\mu+c$, implying there would be a limiting mass of the radial ladder. Physically, this saturation corresponds to string breaking by light-quark pair creation, a consequence of unquenching rather than a property of the quenched flux tube \cite{Bai:2026atm}. Although the confining potential is regular at the origin, we nevertheless smear it with the same Gaussian kernel used for the Coulomb term, in order to treat the finite quark size uniformly. The result is
\begin{widetext}
\begin{equation}
\widetilde{S}_{ij}
  = -\frac{3}{4}C_{ij}\Biggl\{\frac{b\,e^{-\mu r_{ij}}}{4r_{ij}\mu\sigma_{ij}^2}
    \Bigl[4r_{ij}\sigma_{ij}^2 e^{\mu r_{ij}}
    +(\mu-2r_{ij}\sigma_{ij}^2)e^{\mu^2/(4\sigma_{ij}^2)}\mathrm{erfc}\Bigl(\frac{\mu-2r_{ij}\sigma_{ij}^2}{2\sigma_{ij}}\Bigr)
    -(\mu+2r_{ij}\sigma_{ij}^2)e^{\mu^2/(4\sigma_{ij}^2)+2\mu r_{ij}}\mathrm{erfc}\Bigl(\frac{\mu+2r_{ij}\sigma_{ij}^2}{2\sigma_{ij}}\Bigr)\Bigr]
    +c\Biggr\},
\label{equ:Sij}
\end{equation}
\end{widetext}
which we identify with the confining interaction, $V_{ij}^{\rm conf}=\widetilde{S}_{ij}$. Finally, given the confinement kernel (\ref{equ:Sij}), the Thomas-precession term can be written as
\begin{align}
V_{ij}^{\rm so(s)}
  &= -\frac{1}{r_{ij}}\frac{d\widetilde{S}_{ij}}{dr_{ij}}
     \Bigl(\delta_{ii}^{1/2+\epsilon_{\rm so(s)}+\epsilon_{\rm so(s)}^2}\frac{\mathbf{L}\cdot\mathbf{S}_i}{2m_i^2}
     \delta_{ii}^{1/2+\epsilon_{\rm so(s)}+\epsilon_{\rm so(s)}^2} \nonumber\\
     &\qquad
     +\delta_{jj}^{1/2+\epsilon_{\rm so(s)}+\epsilon_{\rm so(s)}^2}\frac{\mathbf{L}\cdot\mathbf{S}_j}{2m_j^2}
     \delta_{jj}^{1/2+\epsilon_{\rm so(s)}+\epsilon_{\rm so(s)}^2}\Bigr).
\label{equ:Vsos}
\end{align}
All spin--angular matrix elements of $\mathbf{S}_i\cdot\mathbf{S}_j$, $\mathbf{L}\cdot\mathbf{S}_i$, $\mathbf{L}\cdot\mathbf{S}_j$, and the tensor operator are evaluated with Clebsch--Gordan coefficients, $6j$, and $9j$ symbols for each $J^P$ channel.

Bound-state masses and wave functions are the eigenvalues and eigenvectors of the Hamiltonian (\ref{equ:GIHamiltonium}). Each meson state is written as a product of color, flavor, spin, and orbital factors:
\begin{equation}
\psi=\psi^{\rm color}\otimes\psi^{\rm flavor}\otimes\psi_{S}^{\rm spin}\otimes\psi_{l}^{\rm orbit},
\label{equ:product}
\end{equation}
where color and flavor are fixed by the $q\bar q'$ configuration. The orbital factor is expanded by the Rayleigh--Ritz method,
\begin{equation}
\psi_{l}^{\rm orbit}=\sum_{n=1}^{N_{\rm max}}C_n\phi_{nl}.
\label{equ:orbit-exp}
\end{equation}
In this work, we employ Gaussian basis functions with a geometric progression~\cite{Hiyama:2003cu,Hiyama:2012sma}, so that a single set covers both the low-lying states and the large spatial extent of radial excitations. In coordinate and momentum space the basis functions read
\begin{equation}
\begin{aligned}
\phi_{nl}(\mathbf{r})
  &= R_{nl}^{r}(r)\,Y_{lm}(\hat{\mathbf{r}})
   = N_{nl}^{r}\,r^{l}\,e^{-\nu_n r^2}Y_{lm}(\hat{\mathbf{r}}),\\
\phi_{nl}(\mathbf{p})
  &= R_{nl}^{p}(p)\,Y_{lm}(\hat{\mathbf{p}})
   = N_{nl}^{p}\,p^{l}\,e^{-p^2/(4\nu_n)}Y_{lm}(\hat{\mathbf{p}}),
\end{aligned}
\label{equ:GEMbasis}
\end{equation}
with normalization constants
\begin{equation}
N_{nl}^{r}
  = \Bigl(\frac{2^{l+2}(2\nu_n)^{l+3/2}}{\sqrt{\pi}(2l+1)!!}\Bigr)^{1/2},\quad
N_{nl}^{p}
  = (-i)^{l}
    \Bigl(\frac{2^{l+2}(2\nu_n)^{-l-3/2}}{\sqrt{\pi}(2l+1)!!}\Bigr)^{1/2}.
\label{equ:GEMnorm}
\end{equation}
The range parameters form a geometric progression,
\begin{equation}
\nu_n=\frac{1}{r_n^2},\quad
r_n=r_{\rm min}\,a^{n-1},\quad
a=\Bigl(\frac{r_{\rm max}}{r_{\rm min}}\Bigr)^{1/(N_{\rm max}-1)},
\label{equ:geom}
\end{equation}
where $r_{\rm min}$ and $r_{\rm max}$ are the smallest and largest length scales retained in the basis, and $N_{\rm max}$ is the number of basis functions.

Since the Gaussians are not mutually orthogonal, the variational problem takes the form of a generalized eigenvalue equation
\begin{equation}
\sum_{n=1}^{N_{\rm max}}\bigl(\langle\phi_{m}|H|\phi_{n}\rangle-E\langle\phi_{m}|\phi_{n}\rangle\bigr)C_n=0.
\label{equ:GEP}
\end{equation}
The Hamiltonian and overlap matrices are
\begin{equation}
\begin{aligned}
H_{mn}&=\langle\phi_{m}|H|\phi_{n}\rangle=\int r^2\,dr\,R_{ml}^{r}\,H\,R_{nl}^{r}, \\
N_{mn}&=\langle\phi_{m}|\phi_{n}\rangle=\int r^2\,dr\,R_{ml}^{r}R_{nl}^{r},
\end{aligned}
\label{equ:HNmat}
\end{equation}
where the angular integral $\int d\Omega\,Y_{l'm'}^{\ast}Y_{lm}=\delta_{l'l}\delta_{m'm}$ has already been handled. Because the Gaussian basis is not orthogonal, the spin-dependent operators cannot be assembled by matrix multiplication in the original set. We therefore first construct an orthogonal basis from the overlap matrix $N$, then evaluate the matrix elements of the spin-dependent interactions. Operators of the form $\hat A(p)\hat B(r)\hat A(p)$ are treated by inserting two complete sets of basis functions,
\begin{align}
&\langle\phi_{m}|\hat A(p)\hat B(r)\hat A(p)|\phi_{n}\rangle \nonumber\\
  &\qquad
  =\sum_{i,j}
    \langle\phi_{m}|\hat A(p)|\phi_{i}\rangle
    \langle\phi_{i}|\hat B(r)|\phi_{j}\rangle
    \langle\phi_{j}|\hat A(p)|\phi_{n}\rangle,
\label{equ:sandwich}
\end{align}
with
\begin{align}
\langle\phi_{m}|\hat A(p)|\phi_{i}\rangle
  &=\int p^2\,dp\,R_{ml}^{p}\,\hat A(p)\,R_{il}^{p},
\label{equ:Ap}\\
\langle\phi_{i}|\hat B(r)|\phi_{j}\rangle
  &=\int r^2\,dr\,R_{il}^{r}\,\hat B(r)\,R_{jl}^{r},
\label{equ:Br}\\
\langle\phi_{j}|\hat A(p)|\phi_{n}\rangle
  &=\int p^2\,dp\,R_{jl}^{p}\,\hat A(p)\,R_{nl}^{p}.
\label{equ:Apn}
\end{align}
The matrix element (\ref{equ:sandwich}) then reduces to independent radial integrals in $r$ and in $p$. With the basis now orthogonal, $H$ is diagonalized as a standard eigenvalue problem. Once the coefficients $C_{ni}$ of the $n$th eigenstate are known for the original Gaussian set, the RMS radius is
\begin{equation}
r_n^{\rm rms}
  =\Biggl(\frac{\sum_{ij}C_{ni}C_{nj}\langle\phi_{il}|r^2|\phi_{jl}\rangle}
               {\sum_{ij}C_{ni}C_{nj}\langle\phi_{il}|\phi_{jl}\rangle}\Biggr)^{1/2}.
\label{equ:rms}
\end{equation}
It serves as an important diagnostic in the termination analysis that follows.

\section{Screening scale from string breaking}
\label{sec:screening}

The radial ladder terminates because the screened confinement (\ref{equ:Vconf}) saturates at a large separation. The corresponding limiting mass appears around $m_i+m_j+b/\mu+c$, indicating a sensitive dependence on the screening parameter $\mu$. A change of a few tens of MeV in $\mu$ moves the limiting mass about hundreds of MeV. Termination is not a prediction until $\mu$ is specified.

For charmonium and bottomonium, $\mu$ is already fixed spectroscopically \cite{Zhang:2026yif}. The high-lying states such as $\psi(4040)$, $\psi(4415)$, $\Upsilon(10860)$, and $\Upsilon(11020)$ constrain the fitted values
\begin{equation}
  \mu_{c\bar c}=0.1440~\mathrm{GeV},\qquad
  \mu_{b\bar b}=0.1222~\mathrm{GeV},
  \label{equ:mu-anchors}
\end{equation}
corresponding to screening lengths $1/\mu_{c\bar c}=1.370~\mathrm{fm}$ and $1/\mu_{b\bar b}=1.615~\mathrm{fm}$. For $D_s$, $B_s$, and $B_c$ the high-lying radial excitations that would control $\mu$ are missing. The low-lying states sit at $r\ll 1/\mu$ and barely feel screening, so a fit of $\mu$ to the $D_s$, $B_s$, and $B_c$ spectrum is degenerate. The only anchor points are therefore the two quarkonium determinations in Eq.~(\ref{equ:mu-anchors}). We estimate the mixed-flavor $\mu$ from string breaking, the unquenching mechanism encoded in the screened confinement.

\subsection{Threshold-dependent string breaking}
\label{sec:energy-balance}

Screening in a single-channel Hamiltonian and coupled-channel dynamics are two descriptions of the same unquenched effect: the creation of a light $n\bar n$ pair ($n=u,d$) that breaks the color string at distance of $1.2$--$1.4$ fm \cite{Bali:2005fu,Castorina:2007eb,Bulava:2019iut,Jiang:2023lmj,Kou:2024dml}. In the coupled-channel picture the $Q\bar Q'$ flux tube does not break at a universal separation. It breaks when the energy stored in the string reaches an open-flavor threshold, where the $Q\bar Q'$ configuration couples to the two-meson continuum~\cite{Luo:2019qkm,Luo:2021dvj,Duan:2021pw,Zhang:2024usz,Man:2024mvl,Man:2025zfu,Man:2025vmm,Qian:2025cc}. Changing the threshold shifts the string-breaking distance in different pair-creation channels \cite{Bulava:2019iut}. A screened potential is an effective encoding of the same physics, so the screening length $1/\mu$ is expected to track the relevant open-flavor threshold of each system. The aim of this subsection is to extract the rule that governs how the distance changes from one system to another.

Concretely, we locate the string-breaking distance of each system from its physical open-flavor threshold. For a static source $Q\bar Q'$ the potential-model energy is $m_Q+m_{\bar Q'}+V(r)$. String breaking into a pair of mesons with masses $M_1$ and $M_2$ is declared at the separation $r_b$ where \cite{Gonzalez:2015hqa}
\begin{equation}
  m_Q+m_{\bar Q'}+V(r_b)
  = M_1+M_2.
  \label{equ:rb-def}
\end{equation}
The sea quark pair is taken to be nonstrange in every case, so the lowest open-flavor channel of each system is
\begin{equation}
\begin{aligned}
  c\bar c &\to D\bar D, &
  b\bar b &\to B\bar B, &
  c\bar s &\to DK, \\
  b\bar s &\to BK, &
  b\bar c &\to B\bar D.
\end{aligned}
\label{equ:channels}
\end{equation}
The meson masses on the right-hand side of Eq.~(\ref{equ:rb-def}) are experimental values~\cite{ParticleDataGroup:2024cfk}, while the quark masses on the left-hand side are the constituent masses of the model.

The screened potential (\ref{equ:Vconf}) already incorporates the unquenched effects through saturation. Therefore, a threshold-dependent analysis of string breaking should be carried out in a purely quenched confinement, where the potential still rises linearly and the open-flavor threshold is imposed from outside rather than built into $V(r)$. We take the Cornell limit $\mu\to 0$,
\begin{equation}
  V(r)
  = \widetilde{G}(r)+br+c,
  \label{equ:Vcornell}
\end{equation}
with the smeared Coulomb kernel $\widetilde{G}$ of Eq.~(\ref{equ:Gij}). All five sources are assigned the unified heavy-meson parameters, so that differences in $r_b$ only arise from the open-flavor thresholds. The constituent masses, $b$, $c$, and the smearing parameters $\sigma_0$ and $s$ are taken from previous work \cite{Zhang:2026yif} as shown in Table~\ref{tab:GIScreen-parameter}, so that the same parameter set keeps the threshold-dependent analysis in the same framework as the spectrum calculation that follows. 

\renewcommand{\tabcolsep}{0.20cm}
\renewcommand{\arraystretch}{1.1}
\begin{table}[!htbp]
\caption{Unified heavy-meson parameters of the screened GI model. The listed $b$ is the unified heavy-meson fit value, without system-dependent specification.}
\label{tab:GIScreen-parameter}
\begin{ruledtabular}
\begin{tabular}{l|l|l}
$m_s=0.6173~{\rm GeV}$ & $c=-0.6482~{\rm GeV}$ & $b=0.2522~{\rm GeV}^2$ \\
$m_c=1.8054~{\rm GeV}$ & $\sigma_0=1.7705~{\rm GeV}$ & \\
$m_b=5.1513~{\rm GeV}$ & $s=1.1463$ & \\
\end{tabular}
\end{ruledtabular}
\end{table}

\begin{figure*}[t]
\centering
\subfigure{\includegraphics[width=0.48\textwidth]{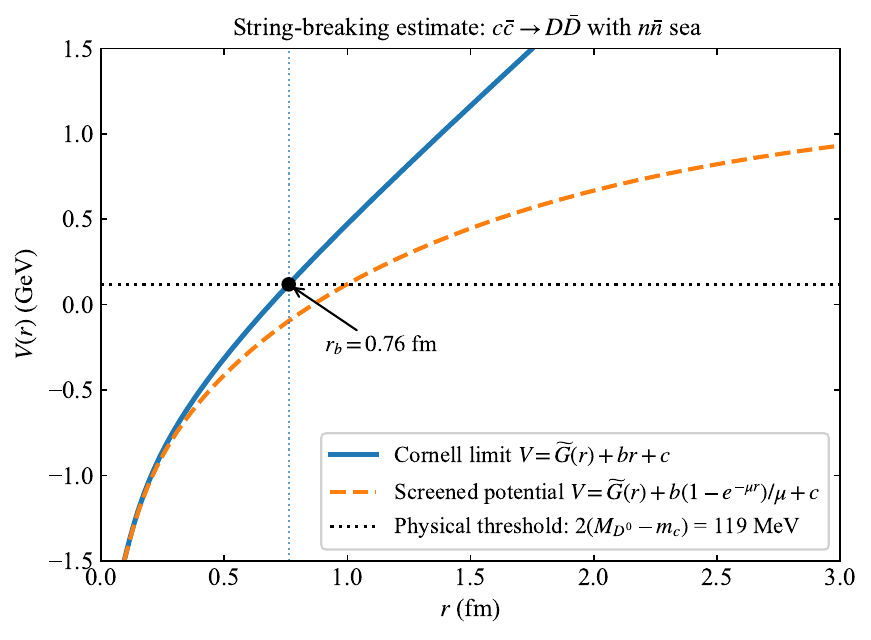}}
\subfigure{\includegraphics[width=0.48\textwidth]{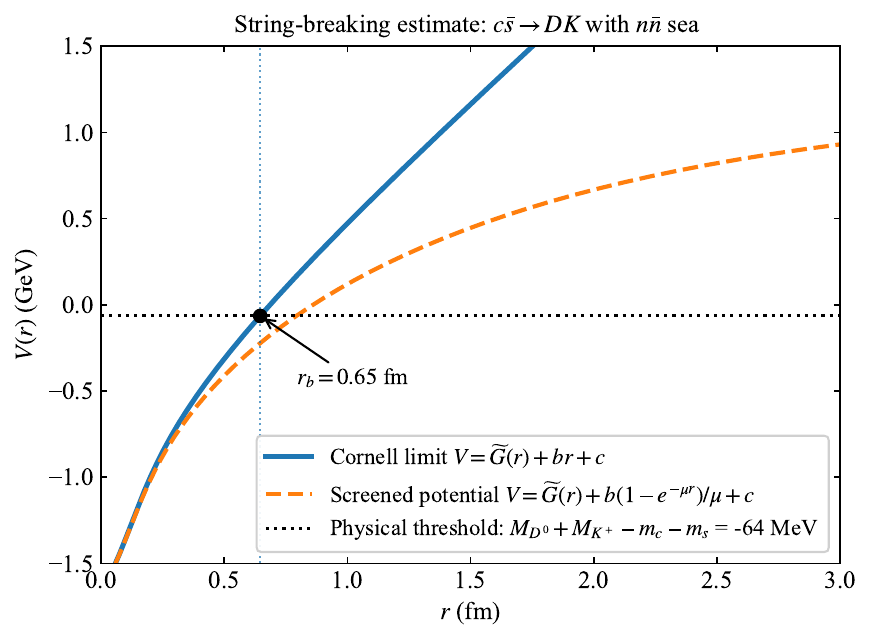}}
\par
\subfigure{\includegraphics[width=0.48\textwidth]{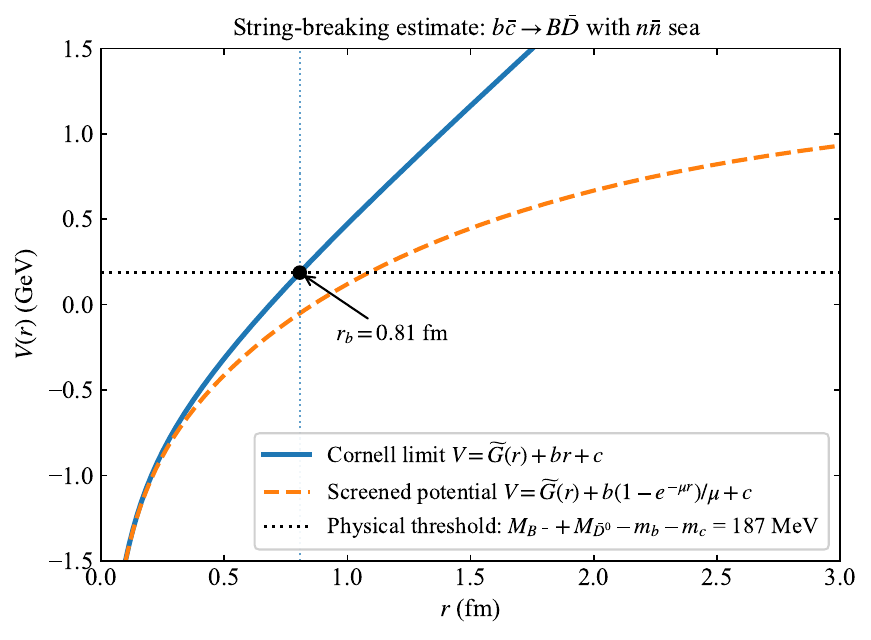}}
\subfigure{\includegraphics[width=0.48\textwidth]{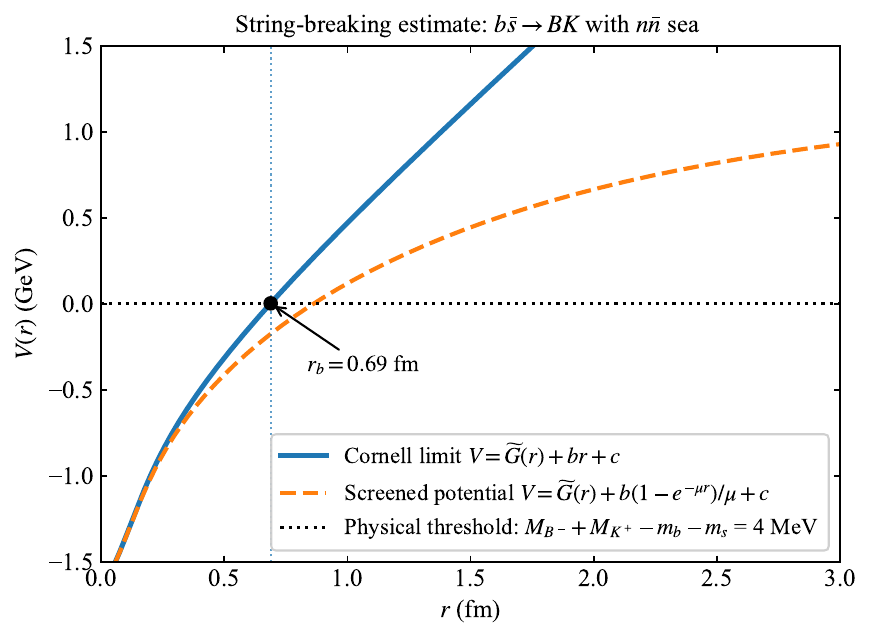}}
\caption{Cornell-limit estimates of $r_b$ for representative sources. The solid curve is the Cornell potential of Eq.~(\ref{equ:Vcornell}); the dashed curve is the screened potential used in this analysis. The dotted horizontal line is the physical open-flavor threshold of Eq.~(\ref{equ:rb-def}). The filled marker locates the estimate $r_b$. }
\label{fig:rb}
\end{figure*}

The resulting distances are collected in Table~\ref{tab:rb} and illustrated in Fig.~\ref{fig:rb}. With a common parameter set the five values are not degenerate: they vary from $0.65~\mathrm{fm}$ in the $DK$ channel to $0.85~\mathrm{fm}$ in the $B\bar B$ channel. The strange systems $c\bar s$ and $b\bar s$ have the lowest open-flavor thresholds and the shortest $r_b$; $b\bar b$ has the highest threshold and the longest $r_b$; $b\bar c$ lies between $c\bar c$ and $b\bar b$. A lower open-flavor threshold is reached sooner, and $r_b$ is shorter. That is the rule extracted here.

\begin{table}[t]
\caption{Cornell-limit estimates of the string-breaking distance $r_b$ from Eq.~(\ref{equ:rb-def}), using the lowest charge-conserving channel of each written source and the unified parameter set. The potential threshold is $V_b=M_1+M_2-m_Q-m_{\bar Q'}$. Replacing a channel by its charged/neutral partner shifts $r_b$ by $\Delta r_b\lesssim 0.006~\mathrm{fm}$.}
\label{tab:rb}
\begin{ruledtabular}
\begin{tabular}{lccccc}
Source & Channel & Partner & $V_b$ (GeV) & $r_b$ (fm) & $\Delta r_b$ (fm) \\ \hline
$c\bar c$ & $D^0\bar D^0$ & $D^+D^-$ & $+0.119$ & $0.762$ & $0.006$ \\
$b\bar b$ & $B^0\bar B^0$ & $B^+B^-$ & $+0.257$ & $0.853$ & $0.000$ \\
$c\bar s$ & $D^0K^+$ & $D^+K^0$ & $-0.064$ & $0.646$ & $0.005$ \\
$b\bar s$ & $B^-K^+$ & $\bar B^0K^0$ & $+0.004$ & $0.689$ & $0.003$ \\
$b\bar c$ & $B^-\bar D^0$ & $\bar B^0 D^-$ & $+0.187$ & $0.807$ & $0.003$ \\
\end{tabular}
\end{ruledtabular}
\end{table}

However, the Cornell limit is not a precision calculation. The constituent masses, $b$, and $c$ were determined in the screened Hamiltonian, not in the original GI model with linear confinement, so taking $\mu\to 0$ directly is inconsistent with the fit. In the $c\bar s$ system it is explicit: the $DK$ threshold lies below $m_c+m_s$, and the potential threshold in Table~\ref{tab:rb} is negative, implying a mismatch between the constituent masses and the physical threshold. Consequently, the values $r_b$ cannot be read as physical string-breaking distances. The pattern among them still supplies the quantitative relation from one system to another, which is converted to the values of $\mu$ in Sec.~\ref{sec:mu-mixed}.

\subsection{Screening parameters for mixed-flavor systems}
\label{sec:mu-mixed}

On the two spectroscopic anchors of Eq.~(\ref{equ:mu-anchors}), $\mu$ is known from the previous fit, and the screening length lies beyond the Cornell-limit $r_b$:
\begin{equation}
\begin{aligned}
  \Delta_{c\bar c}
  &= 1/\mu_{c\bar c}-r_b(c\bar c)
   = 0.609~\mathrm{fm},\\
  \Delta_{b\bar b}
  &= 1/\mu_{b\bar b}-r_b(b\bar b)
   = 0.762~\mathrm{fm}.
\end{aligned}
\label{equ:Delta-anchors}
\end{equation}
The offset $\Delta$ converts the estimated $r_b$ into the screening length $1/\mu$ that enters the heavy-meson spectrum. We interpolate $\Delta$ in the reduced mass of the static source,
\begin{equation}
  t
  =\frac{m_{\rm red}-m_{\rm red}^{c\bar c}}
        {m_{\rm red}^{b\bar b}-m_{\rm red}^{c\bar c}},
  \qquad
  m_{\rm red}=\frac{m_Q m_{\bar Q'}}{m_Q+m_{\bar Q'}},
  \label{equ:t-red}
\end{equation}
and, with the additive offset, convert the pattern of Table~\ref{tab:rb} into the central values of $\mu$:
\begin{equation}
  \frac{1}{\mu}
  = r_b+\Delta_{c\bar c}+t\bigl(\Delta_{b\bar b}-\Delta_{c\bar c}\bigr).
  \label{equ:form1}
\end{equation}
If every $r_b$ is shifted by the same additive constant, $\Delta$ at the two anchors changes by the opposite amount, and $1/\mu$ in Eq.~(\ref{equ:form1}) remains unchanged. Small corrections of this kind---the choice of charged/neutral open-flavor partner ($\Delta r_b\lesssim 0.006~\mathrm{fm}$) and the PDG mass errors---are therefore not treated as uncertainties on $\mu$. 

The conversion from the pattern of $r_b$ to $\mu$ is not unique. A ratio $1/\mu=k(m_{\rm red})\,r_b$ multiplies the mismatch in $r_b$ and is discarded. The remaining conversions differ, and that spread is the window on $\mu$:
\begin{enumerate}[(i)]
\item the additive offset of Eq.~(\ref{equ:form1}). It supplies the central values $\mu(D_s)=0.1626~\mathrm{GeV}$, $\mu(B_s)=0.1560~\mathrm{GeV}$, and $\mu(B_c)=0.1356~\mathrm{GeV}$. For $D_s$ this is the upper edge of the window. For $B_c$, where $t$ lies between the two anchors, it lies between (ii) and (iii).
\item the linear conversion $1/\mu=\alpha+\beta r_b$ through the two anchors, with $\beta=(1/\mu_{b\bar b}-1/\mu_{c\bar c})/\bigl(r_b(b\bar b)-r_b(c\bar c)\bigr)$. For $B_c$ it agrees with (i) to a few MeV and gives the lower edge of the window. For $B_s$, where $t$ lies below the charmonium anchor, it gives the upper edge. For $D_s$ it yields $\mu=0.186~\mathrm{GeV}$, whose screening length lies below the lattice scale \cite{Bali:2005fu,Bulava:2019iut}, and is omitted from the window.
\item the interpolation of $\mu$ in reduced mass, $\mu=\mu_{c\bar c}+t\bigl(\mu_{b\bar b}-\mu_{c\bar c}\bigr)$, without using $r_b$. It gives the lower edges of the $D_s$ and $B_s$ windows, $0.150~\mathrm{GeV}$ and $0.149~\mathrm{GeV}$. For $B_c$ it gives upper edge $\mu=0.138~\mathrm{GeV}$, close to (i).
\end{enumerate}
The resulting central values and windows are listed in Table~\ref{tab:mu-mixed}. The central values are the screening parameters used in the $D_s$, $B_s$, and $B_c$ spectra, not spectroscopic fits of those systems. The quoted ranges reflect the spread among alternative conversions, not a statistical uncertainty. The spectrum calculation that follows uses the central values of Table~\ref{tab:mu-mixed} and varies $\mu$ across these windows, with the remaining parameters held fixed. That variation is the uncertainty on the termination of the radial ladder.

\begin{table}[t]
\caption{Central screening parameters for mixed-flavor heavy mesons from the additive offset (\ref{equ:form1}), with conversion windows on $\mu$. The charmonium and bottomonium rows are the spectroscopic anchors of Eq.~(\ref{equ:mu-anchors}), listed for comparison.}
\label{tab:mu-mixed}
\begin{ruledtabular}
\begin{tabular}{lcccc}
System & $m_{\rm red}$ (GeV) & $1/\mu$ (fm) & $\mu$ (GeV) & Window (GeV) \\ \hline
$c\bar c$ & $0.903$ & $1.370$ & $0.1440$ & --- \\
$b\bar b$ & $2.576$ & $1.615$ & $0.1222$ & --- \\
$D_s$ ($c\bar s$) & $0.460$ & $1.214$ & $0.1626$ & $0.150$--$0.163$ \\
$B_s$ ($b\bar s$) & $0.551$ & $1.265$ & $0.1560$ & $0.149$--$0.168$ \\
$B_c$ ($b\bar c$) & $1.337$ & $1.455$ & $0.1356$ & $0.132$--$0.138$ \\
\end{tabular}
\end{ruledtabular}
\end{table}

\section{Numerical results}
\label{sec:results}

The screening parameters of Table~\ref{tab:mu-mixed} are held fixed at their central values. The remaining system-dependent parameters are then determined from the established $D_s$, $B_s$, and $B_c$ states, and the full radial ladders are computed in the same Gaussian basis as in Sec.~\ref{sec:framework}. Subsection~\ref{sec:spectra} reports the fits and the spectra. Subsection~\ref{sec:termination} locates the termination of each radial ladder.

\subsection{Fitted parameters and spectra of \texorpdfstring{$D_s$, $B_s$, and $B_c$}{Ds, Bs, and Bc}}
\label{sec:spectra}

Quark masses, the constant $c$, and the smearing parameters $\sigma_0$ and $s$ are taken from the unified heavy-meson solution of Table~\ref{tab:GIScreen-parameter}. The screening parameter $\mu$ of each mixed-flavor system is the central value of Table~\ref{tab:mu-mixed}. The Coulomb exponent is kept at the GI default, $\epsilon_{\rm Coul}=0$. With those quantities fixed, five parameters remain free in each family: the string tension $b$ and the relativistic exponents $\epsilon_{\rm cont}$, $\epsilon_{\rm so(\nu)}$, $\epsilon_{\rm so(s)}$, and $\epsilon_{\rm tens}$. The $\chi^2$ function is
\begin{equation}
  \chi^{2}
  = \sum_{i}
    \left(\frac{M_{i}^{\rm th}-M_{i}^{\rm exp}}{\sigma_{i}}\right)^{2},
  \label{equ:chi2}
\end{equation}
where $M_{i}^{\rm th}$ is the theoretical evaluation, $M_{i}^{\rm exp}$ is the experimental anchor and $\sigma_{i}$ is the adopted error. Minimization is performed with ROOT:Minuit2 and Migrad algorithm \cite{ROOT_NIMA_1997}. Only states with a unique spectroscopic assignment enter Eq.~(\ref{equ:chi2}). The resulting parameters are collected in Table~\ref{tab:fit-mixed}. In every family $b$ barely shifts from the unified heavy-meson value $0.2522~{\rm GeV}^{2}$ used for the string-breaking analysis of Sec.~\ref{sec:screening}. The contact exponent is constrained by the $S$-wave hyperfine splitting among families. The spin--orbit and tensor exponents are determined precisely only in $D_s$. In $B_s$ and $B_c$ those exponents remain unconstrained: four and three anchors, respectively, cannot determine five free parameters.

\renewcommand{\tabcolsep}{0.12cm}
\renewcommand{\arraystretch}{1.1}
\begin{table}[t]
\caption{System-dependent parameters of the screened GI Hamiltonian for mixed-flavor heavy mesons. Quark masses, $c$, $\sigma_0$, and $s$ are the unified values of Table~\ref{tab:GIScreen-parameter}. Fixed quantities are marked with an asterisk. $N$ is the number of $\chi^{2}$ anchors.}
\label{tab:fit-mixed}
\begin{ruledtabular}
\begin{tabular}{lccc}
 & $D_s$ & $B_s$ & $B_c$ \\ \hline
$N$ & $8$ & $4$ & $3$ \\
$\chi^{2}$ & $1700$ & $159.5$ & $8.97$ \\
$b$ (${\rm GeV}^{2}$) & $0.2576$ & $0.2595$ & $0.2549$ \\
$\mu$ (${\rm GeV}$) & $0.1626^{\ast}$ & $0.1560^{\ast}$ & $0.1356^{\ast}$ \\
$\epsilon_{\rm Coul}$ & $0^{\ast}$ & $0^{\ast}$ & $0^{\ast}$ \\
$\epsilon_{\rm cont}$ & $-0.280$ & $-0.320$ & $-0.499$ \\
$\epsilon_{\rm so(\nu)}$ & $-0.500$ & $-0.295$ & $-0.343$ \\
$\epsilon_{\rm so(s)}$ & $1.000$ & $-0.169$ & $0.791$ \\
$\epsilon_{\rm tens}$ & $-0.500$ & $-1.000$ & $-0.500$ \\
\end{tabular}
\end{ruledtabular}
\end{table}

\renewcommand{\tabcolsep}{0.18cm}
\renewcommand{\arraystretch}{1.1}
\begin{table}[t]
\caption{Calculated masses compared with the experimental $\chi^{2}$ anchors. Only states that enter Eq.~(\ref{equ:chi2}) are listed. Masses and $\sigma$ are in MeV. Experimental values are those used in the fits.}
\label{tab:fit-anchors}
\begin{ruledtabular}
\begin{tabular}{lccrrl}
$n^{2S+1}L_{J}$ & $J^{P}$ & $M_{\rm th}$ & $M_{\rm exp}$ & $\sigma$ & State \\ \hline
\multicolumn{6}{l}{$D_s$} \\
$1^{1}S_{0}$ & $0^{-}$ & $1971.9$ & $1968.4$ & $1$ & $D_{s}$ \\
$1^{3}S_{1}$ & $1^{-}$ & $2103.2$ & $2112.2$ & $1$ & $D_{s}^{\ast}$ \\
$1^{3}P_{0}$ & $0^{+}$ & $2340.8$ & $2317.8$ & $1$ & $D_{s0}^{\ast}(2317)$ \\
$1^{3}P_{1}$ & $1^{+}$ & $2474.6$ & $2459.5$ & $1$ & $D_{s1}(2460)$ \\
$1^{1}P_{1}$ & $1^{+}$ & $2506.1$ & $2535.1$ & $1$ & $D_{s1}(2536)$ \\
$1^{3}P_{2}$ & $2^{+}$ & $2568.4$ & $2569.1$ & $1$ & $D_{s2}^{\ast}(2573)$ \\
$2^{1}S_{0}$ & $0^{-}$ & $2600.9$ & $2591.0$ & $9$ & $D_{s0}(2590)$ \\
$1^{3}D_{3}$ & $3^{-}$ & $2839.8$ & $2860.5$ & $7$ & $D_{s3}^{\ast}(2860)$ \\
\multicolumn{6}{l}{$B_s$} \\
$1^{1}S_{0}$ & $0^{-}$ & $5370.3$ & $5366.9$ & $1$ & $B_{s}$ \\
$1^{3}S_{1}$ & $1^{-}$ & $5424.2$ & $5415.4$ & $1$ & $B_{s}^{\ast}$ \\
$1^{1}P_{1}$ & $1^{+}$ & $5820.3$ & $5828.7$ & $1$ & $B_{s1}(5830)$ \\
$1^{3}P_{2}$ & $2^{+}$ & $5839.9$ & $5839.9$ & $1$ & $B_{s2}^{\ast}(5840)$ \\
\multicolumn{6}{l}{$B_c$} \\
$1^{1}S_{0}$ & $0^{-}$ & $6276.9$ & $6274.5$ & $1$ & $B_{c}$ \\
$1^{3}S_{1}$ & $1^{-}$ & $6341.0$ & $6339.0$ & $2$ & $B_{c}^{\ast}$ \\
$2^{1}S_{0}$ & $0^{-}$ & $6869.7$ & $6871.2$ & $1$ & $B_{c}(2S)$ \\
\end{tabular}
\end{ruledtabular}
\end{table}

The $\chi^{2}$ anchors of all three families are collected in Table~\ref{tab:fit-anchors}. For $D_s$, eight established levels are fitted. Every $1S$ and $1P$ anchor carries a $1~{\rm MeV}$ error; $D_{s0}(2590)$ is inflated to $9~{\rm MeV}$, while $D_{s3}^{\ast}(2860)$ keeps $7~{\rm MeV}$ experimental error. The vector pair $D_{s1}^{\ast}(2700)$ and $D_{s1}^{\ast}(2860)$ is excluded as a $2S$--$1D$ mixed pair~\cite{Song:2015nia}, not as pure $2^{3}S_{1}$ and $1^{3}D_{1}$ eigenstates. $D_{s1}(2933)$ is not assigned and does not enter the fit~\cite{LHCb:2026ds2933}. The $1S$ masses and $D_{s2}^{\ast}(2573)$ are reproduced at the few-MeV level, $D_{s0}(2590)$ to $10~{\rm MeV}$, and the $1^{3}D_{3}$ lies $21~{\rm MeV}$ below $D_{s3}^{\ast}(2860)$. The calculated $1^{3}P_{0}$ and $1^{3}P_{1}$ lie $23$ and $15~{\rm MeV}$ above $D_{s0}^{\ast}(2317)$ and $D_{s1}(2460)$; the $1^{1}P_{1}$ assignment of $D_{s1}(2536)$ is $29~{\rm MeV}$ low. The resulting $\chi^{2}=1700$ is a model-precision and coupled-channel, not a failure of the minimizer. For $B_s$, four established states are fitted. The four anchors are reproduced to better than $9~{\rm MeV}$, with $\chi^{2}=159.5$ set by the $1~{\rm MeV}$ error. For $B_c$, three $S$-wave anchors are fitted. Three anchors cannot determine five free parameters; the fit constrains $b$ and the hyperfine splitting, not the spin--orbit or tensor exponents. Residuals on the three anchors are $2$, $2$, and $-1$~MeV, with $\chi^{2}=8.97$.

The $D_s$ and $B_s$ spectra for $n=1$--$10$ and the $B_c$ spectrum for $n=1$--$14$ are listed in Appendix~\ref{app:spectra} and displayed in Fig.~\ref{fig:spectra}. Every family develops a mass plateau, while the RMS radii continue to grow rapidly. That mass-radius decoupling is the spectroscopic signature of a screened confinement, and it is the foundation of the termination analysis in Sec.~\ref{sec:termination}.

\begin{figure*}[t]
\centering
\includegraphics[width=0.32\textwidth]{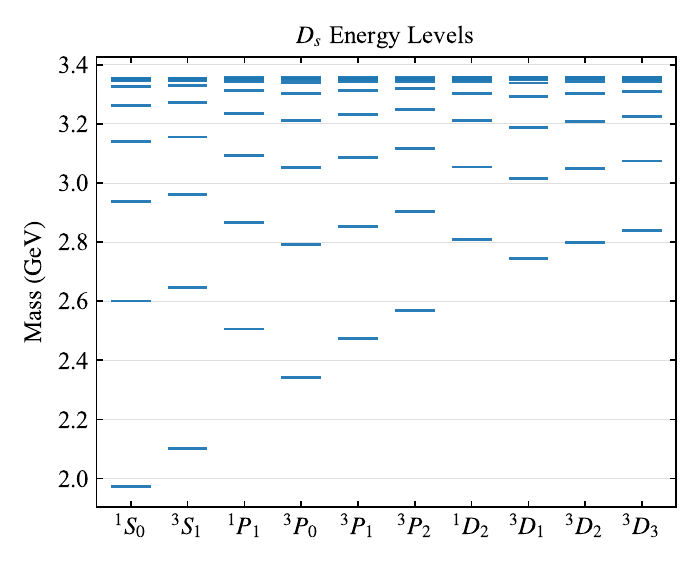}\hfill
\includegraphics[width=0.32\textwidth]{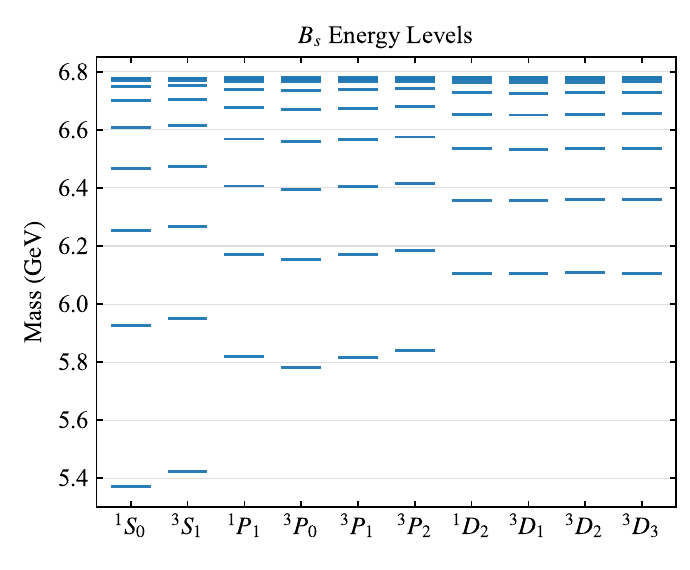}\hfill
\includegraphics[width=0.32\textwidth]{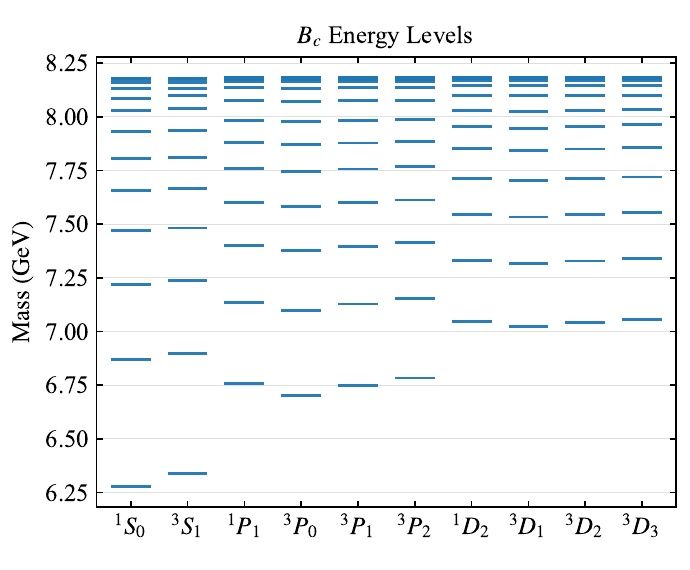}
\par
\includegraphics[width=0.32\textwidth]{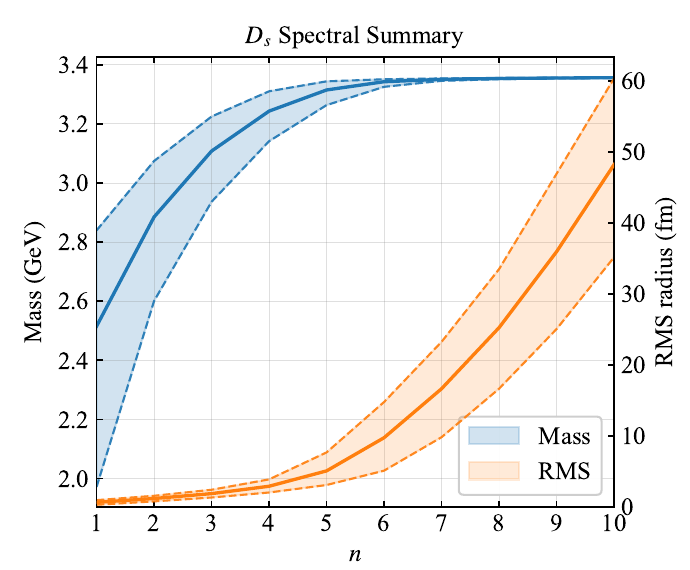}\hfill
\includegraphics[width=0.32\textwidth]{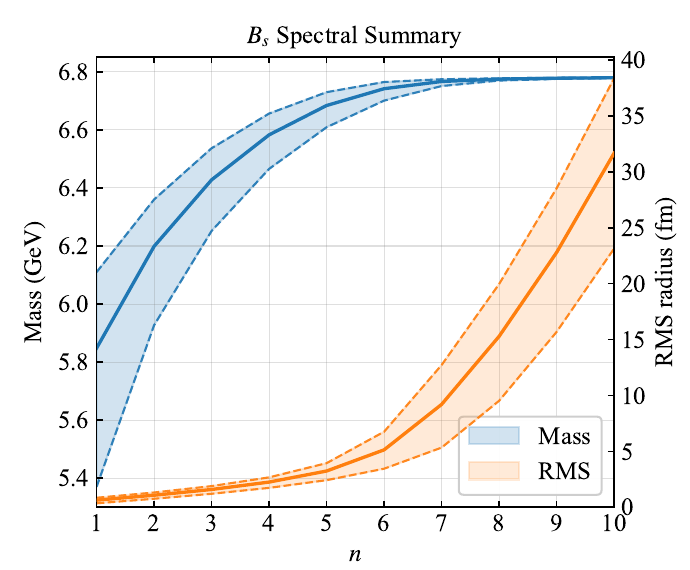}\hfill
\includegraphics[width=0.32\textwidth]{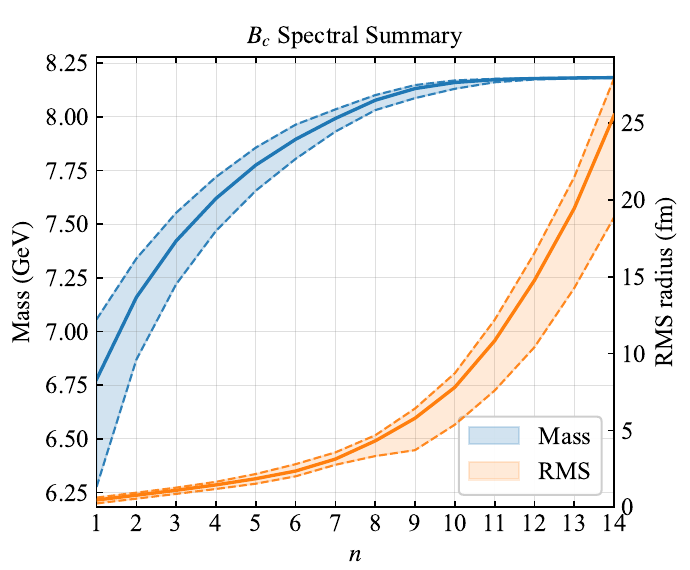}
\caption{Energy levels (upper row) and spectral summaries (lower row) of $D_s$, $B_s$, and $B_c$ families. Compression of neighboring levels at the top of each column is the onset of the mass plateau. In the summaries, the blue (orange) band is the range of mass (RMS radius) over the ten $S$, $P$, and $D$ channels, and the solid curve is the channel average.}
\label{fig:spectra}
\end{figure*}

\subsection{Termination of the radial ladder}
\label{sec:termination}

As suggested by string-breaking, the screened confinement saturates at large separation, so the radial ladder cannot continue infinitely. The asymptotic mass implied by Eq.~(\ref{equ:Vconf}) is
\begin{equation}
  M_{\infty}
  = m_{i}+m_{j}+\frac{b}{\mu}+c.
  \label{equ:Mlim}
\end{equation}
With the parameters of Table~\ref{tab:fit-mixed} this limiting mass is
\begin{equation}
\begin{aligned}
  M_{\infty}(D_s)&=3.359~{\rm GeV},\\
  M_{\infty}(B_s)&=6.784~{\rm GeV},\\
  M_{\infty}(B_c)&=8.188~{\rm GeV}.
\end{aligned}
\label{equ:Mlim-num}
\end{equation}
The extended spectra of Appendix~\ref{app:spectra} and Fig.~\ref{fig:spectra} show that every channel approaches the corresponding plateau, while the RMS radius continues to increase. That is the mass-radius decoupling addressed in this work. A state remains physically interpretable only while neighboring levels are energetically resolvable and spatially compact. The primary quantities are therefore the mass gaps and RMS-radius increments,
\begin{equation}
  \Delta M_{n}=M_{n}-M_{n-1},
  \label{equ:dM}
\end{equation}
\begin{equation}
  \Delta R_{n}=R_{n}-R_{n-1},
  \label{equ:dR}
\end{equation}
with $\Delta M_{n}$ in MeV and $\Delta R_{n}$ in fm. The termination boundary is the onset of this mass-radius decoupling.

This boundary is identified when the next two radial levels have mass gaps below $10~{\rm MeV}$ while the RMS radius still grows by several fm. For $D_s$ and $B_s$ the threshold is
\begin{equation}
\begin{aligned}
  &\min(\Delta M_{n+1},\Delta M_{n+2})<10~{\rm MeV}, \\
  &\max(\Delta R_{n+1},\Delta R_{n+2})>4~{\rm fm}.
\end{aligned}
\label{equ:caution-ds}
\end{equation}
For $B_c$ it is
\begin{equation}
\begin{aligned}
  &\min(\Delta M_{n+1},\Delta M_{n+2})<10~{\rm MeV}, \\
  &\max(\Delta R_{n+1},\Delta R_{n+2})>2~{\rm fm},
\end{aligned}
\label{equ:caution-bsbc}
\end{equation}
inspired from the previous analysis \cite{Zhang:2026yif}. Let $C_{n}$ denote Eq.~(\ref{equ:caution-ds}) or Eq.~(\ref{equ:caution-bsbc}), according to the family. Set $S=\{1,2,\dots,10\}$ for $D_s$ and $B_s$, and $S=\{1,2,\dots,14\}$ for $B_c$. The first index in $S$ that satisfies $C_{n}$ is the caution onset,
\begin{equation}
  n_{\mathrm{caution}}
  =\min\bigl\{n\in S:C_{n}\bigr\}.
  \label{equ:ncaution}
\end{equation}
The safe range therefore ends at
\begin{equation}
  n_{\mathrm{safe}}=n_{\mathrm{caution}}-1.
  \label{equ:nsafe}
\end{equation}
The resulting channel-by-channel boundaries are summarized in Table~\ref{tab:termination} of Appendix~\ref{app:spectra}. The $D_s$ ladder terminates first: $n_{\mathrm{safe}}=6$ in $S$ waves, $5$ in $P$ waves, and $4$ in most $D$ waves, with the plateau already reached near $3.35~{\rm GeV}$. The $B_s$ ladder, using the same threshold as $D_s$, extends one unit further, $n_{\mathrm{safe}}=7$ ($S$), $6$ ($P$), and $5$--$6$ ($D$), and saturates near $6.78~{\rm GeV}$. The $B_c$ ladder is the longest, $n_{\mathrm{safe}}=11$ ($S$), $10$ ($P$), and $9$ ($D$), and saturates near $8.18~{\rm GeV}$. In every family the $D$ waves reach the termination boundary first, followed by the $P$ and $S$ waves, as expected from the larger spatial extent of higher orbital angular momentum.

This ordering follows from the screening lengths of Table~\ref{tab:mu-mixed}. The $D_s$ system has the largest $\mu$ and the lightest reduced mass, so the confining potential saturates earliest and the RMS radii inflate after only a few radial nodes. The $B_c$ system has the smallest $\mu$ and a compact $b\bar c$ wave function, so many radial excitations remain distinguishable before $\Delta M_{n}$ collapses.

Supporting changepoint tests are constructed from the spectrum alone. Define the indicator
\begin{equation}
  f_{n}=\log_{10}\!\left(\frac{\Delta R_{n}+\epsilon_{R}}{\Delta M_{n}+\epsilon_{M}}\right),
  \label{equ:fn}
\end{equation}
where $\Delta M_{n}$ is in MeV, $\Delta R_{n}$ is in fm, and $\epsilon_{R}=0.001~{\rm fm}$, $\epsilon_{M}=0.001~{\rm MeV}$ are introduced only to avoid numerical instability. The quantity $f_{n}$ increases when the RMS radius grows while the mass compresses. After standardizing the sequence,
\begin{equation}
  z_{n}=\frac{f_{n}-\bar f}{\sigma_{f}},
  \label{equ:zn}
\end{equation}
with $\bar f$ and $\sigma_{f}$ the mean and standard deviation of $\{f_{n}\}$, the cumulative-sum (CUSUM) process is
\begin{equation}
  S_{k}=\sum_{n\le k}z_{n}.
  \label{equ:Sk}
\end{equation}
If the sequence is homogeneous, $S_{k}$ remains balanced. A transition from an early regime to a later one produces a systematic drift, and the corresponding changepoint is
\begin{equation}
  n_{\mathrm{CUSUM}}=\operatorname*{arg\,max}_{k}\lvert S_{k}\rvert.
  \label{equ:nCUSUM}
\end{equation}
As a second test, the sequence is split at a candidate index $j$ into early and late segments. The residual sum of squares is
\begin{equation}
  \mathrm{RSS}(j)
  =\sum_{n<j}\bigl(f_{n}-\bar f_{<j}\bigr)^{2}
  +\sum_{n\ge j}\bigl(f_{n}-\bar f_{\ge j}\bigr)^{2},
  \label{equ:RSS}
\end{equation}
where $\bar f_{<j}$ and $\bar f_{\ge j}$ are the means of the two segments. The two-segment changepoint is
\begin{equation}
  n_{\mathrm{2\text{-}seg}}=\operatorname*{arg\,min}_{j}\mathrm{RSS}(j).
  \label{equ:n2seg}
\end{equation}
CUSUM locates the largest cumulative drift. The two-segment split locates the best two-regime description. Both tests are supporting only and do not override Eqs.~(\ref{equ:ncaution}) and~(\ref{equ:nsafe}).

A further supporting quantity is the dimensionless resolvability
\begin{equation}
  \chi_{n}=\Delta M_{n}\,r_{n},
  \label{equ:chi-n}
\end{equation}
with $\Delta M_{n}$ in GeV and $r_{n}$ converted to ${\rm GeV}^{-1}$. Collapse of $\chi_{n}$ below unity marks the loss of spectral identifiability. The first such index, $n_{\chi<1}$, and the corresponding value $\chi_{\chi<1}$ are listed with the changepoints in Table~\ref{tab:termination}. They follow the same pattern as the primary threshold: $\chi_{n}$ falls below unity near $n=6$--$8$ in $D_s$, $n=7$--$8$ in $B_s$, and $n=10$--$12$ in $B_c$.

Finally, the conversion windows on $\mu$ in Table~\ref{tab:mu-mixed} shift $M_{\infty}$ through $b/\mu$ at fixed $b$. The induced intervals are
\begin{equation}
\begin{aligned}
  M_{\infty}(D_s) &= 3.36\text{--}3.49~{\rm GeV},\\
  M_{\infty}(B_s) &= 6.66\text{--}6.86~{\rm GeV},\\
  M_{\infty}(B_c) &= 8.16\text{--}8.24~{\rm GeV}.
\end{aligned}
\label{equ:Mlim-window}
\end{equation}
A $30$--$50~{\rm MeV}$ movement of the $B_c$ plateau is of the same order as the last safe $S$-wave gaps and does not change $n_{\rm safe}$. A $130~{\rm MeV}$ upward shift of the $D_s$ plateau, corresponding to the lower edge $\mu=0.150~{\rm GeV}$, is comparable to the $n=4$--$5$ gap and may move $n_{\rm safe}$ by one unit. Within every window the qualitative conclusion is unchanged: the mixed-flavor radial ladder terminates, first in $D_s$, then in $B_s$, and last in $B_c$.

\section{Summary and Conclusions}
\label{sec:summary}

In this work the spectra of the mixed-flavor heavy mesons $D_s$ ($c\bar s$), $B_s$ ($b\bar s$), and $B_c$ ($b\bar c$) are obtained from a screen-modified Godfrey--Isgur Hamiltonian, solved with the Gaussian expansion method. The screened confinement replaces the linear potential and saturates at large separation, so each radial ladder has a finite limiting mass $M_{\infty}=m_i+m_j+b/\mu+c$. The calculated spectra reach this mass, while the root-mean-square radii continue to grow rapidly. Once neighboring levels are no longer resolvable in mass and the spatial extent is no longer compact, the ladder comes to an end. This mass-radius decoupling is the signature of the termination, and its location is controlled by the screening length.

The screening length is fixed by string breaking through light-quark pair creation. It cannot be taken from a spectroscopic fit, because the excitations that would determine it have not been observed. For each system the string-breaking distance is set by the lowest open-flavor threshold, $DK$, $BK$, or $B\bar D$. A lower threshold is reached at a shorter distance and corresponds to a stronger screening. Matching these distances to the charmonium and bottomonium anchors gives $\mu(D_s)=0.1626~{\rm GeV}$, $\mu(B_s)=0.1560~{\rm GeV}$, and $\mu(B_c)=0.1356~{\rm GeV}$. The termination is then a prediction of the unquenched confinement, with $D_s$ screened most strongly and $B_c$ least strongly. The uncertainty on that prediction is the conversion window on each value of $\mu$.

The calculated ladders follow this order. The $D_s$ ladder ends at $n_{\mathrm{safe}}=6$, $5$, and $4$ in the $S$, $P$, and most $D$ waves, near $3.4~{\rm GeV}$. The $B_s$ ladder ends one step higher, near $6.8~{\rm GeV}$. The $B_c$ ladder ends last, at $n_{\mathrm{safe}}=11$, $10$, and $9$, near $8.2~{\rm GeV}$. Within each family the $D$ waves terminate before the $P$ waves, and the $P$ waves before the $S$ waves. Across the uncertainty in $\mu$, the $D_s$ boundary may move by one unit. The $B_c$ safe range remains fixed, and the order of the three families is unchanged. This termination is directly testable with forthcoming high-statistics data from BESIII, Belle II, and LHCb. Above the upper edges of the limiting-mass windows, $3.49~{\rm GeV}$ ($D_s$), $6.86~{\rm GeV}$ ($B_s$), and $8.24~{\rm GeV}$ ($B_c$), states of the radial ladder cannot be discovered.

\appendix
\section{Extended spectra and termination diagnostics}
\label{app:spectra}

Tables~\ref{tab:Ds-spectra} and~\ref{tab:Bs-spectra} list the $D_s$ and $B_s$ masses and RMS radii for $n=1$--$10$. Table~\ref{tab:Bc-spectra} lists the $B_c$ spectrum through $n=14$. Table~\ref{tab:termination} records the termination indices defined in Sec.~\ref{sec:termination}: the threshold-based pair $(n_{\mathrm{safe}},n_{\mathrm{caution}})$, the CUSUM and two-segment changepoints of $f_{n}$, and the first collapse of $\chi_{n}$ below unity.

\renewcommand{\tabcolsep}{0.11cm}
\renewcommand{\arraystretch}{1.05}
\begin{table*}[t]
\caption{Extended $D_s$ spectrum for the $nS$, $nP$, and $nD$ families with $n=1$--$10$. For each channel the calculated mass $M$ (GeV) and RMS radius $r$ (fm) are listed.}
\label{tab:Ds-spectra}
\begin{ruledtabular}
\begin{tabular}{cl*{10}{c}}
$^{2S+1}L_{J}$ & & \multicolumn{10}{c}{$n$} \\
 & & $1$ & $2$ & $3$ & $4$ & $5$ & $6$ & $7$ & $8$ & $9$ & $10$ \\ \hline
$^{1}S_{0}$ & $M$ & $1.972$ & $2.601$ & $2.937$ & $3.142$ & $3.264$ & $3.326$ & $3.346$ & $3.351$ & $3.354$ & $3.355$ \\
 & $r$ & $0.35$ & $0.83$ & $1.38$ & $2.08$ & $3.14$ & $5.16$ & $9.87$ & $16.69$ & $25.09$ & $35.20$ \\
$^{3}S_{1}$ & $M$ & $2.103$ & $2.647$ & $2.961$ & $3.156$ & $3.271$ & $3.329$ & $3.347$ & $3.352$ & $3.354$ & $3.356$ \\
 & $r$ & $0.41$ & $0.88$ & $1.43$ & $2.16$ & $3.26$ & $5.43$ & $10.40$ & $17.34$ & $25.89$ & $36.31$ \\
$^{1}P_{1}$ & $M$ & $2.506$ & $2.866$ & $3.094$ & $3.236$ & $3.314$ & $3.343$ & $3.351$ & $3.354$ & $3.355$ & $3.356$ \\
 & $r$ & $0.64$ & $1.16$ & $1.80$ & $2.71$ & $4.31$ & $8.22$ & $14.85$ & $22.99$ & $32.91$ & $46.07$ \\
$^{3}P_{0}$ & $M$ & $2.341$ & $2.791$ & $3.052$ & $3.212$ & $3.302$ & $3.340$ & $3.350$ & $3.353$ & $3.355$ & $3.356$ \\
 & $r$ & $0.52$ & $1.04$ & $1.65$ & $2.49$ & $3.87$ & $7.09$ & $13.35$ & $21.15$ & $30.71$ & $43.24$ \\
$^{3}P_{1}$ & $M$ & $2.475$ & $2.851$ & $3.086$ & $3.231$ & $3.312$ & $3.343$ & $3.350$ & $3.354$ & $3.355$ & $3.356$ \\
 & $r$ & $0.62$ & $1.13$ & $1.77$ & $2.66$ & $4.21$ & $7.96$ & $14.52$ & $22.59$ & $32.44$ & $45.59$ \\
$^{3}P_{2}$ & $M$ & $2.568$ & $2.902$ & $3.117$ & $3.249$ & $3.320$ & $3.345$ & $3.351$ & $3.354$ & $3.355$ & $3.356$ \\
 & $r$ & $0.71$ & $1.24$ & $1.90$ & $2.86$ & $4.63$ & $9.00$ & $15.83$ & $24.19$ & $34.26$ & $46.97$ \\
$^{1}D_{2}$ & $M$ & $2.808$ & $3.054$ & $3.212$ & $3.304$ & $3.342$ & $3.351$ & $3.354$ & $3.355$ & $3.356$ & $3.357$ \\
 & $r$ & $0.94$ & $1.54$ & $2.36$ & $3.71$ & $7.08$ & $13.97$ & $22.26$ & $32.28$ & $45.11$ & $57.94$ \\
$^{3}D_{1}$ & $M$ & $2.744$ & $3.016$ & $3.189$ & $3.292$ & $3.339$ & $3.350$ & $3.353$ & $3.355$ & $3.356$ & $3.357$ \\
 & $r$ & $0.83$ & $1.42$ & $2.18$ & $3.39$ & $6.14$ & $12.59$ & $20.59$ & $30.23$ & $41.89$ & $53.55$ \\
$^{3}D_{2}$ & $M$ & $2.798$ & $3.048$ & $3.209$ & $3.302$ & $3.342$ & $3.351$ & $3.354$ & $3.355$ & $3.356$ & $3.357$ \\
 & $r$ & $0.92$ & $1.52$ & $2.32$ & $3.65$ & $6.91$ & $13.74$ & $21.98$ & $31.93$ & $44.56$ & $57.19$ \\
$^{3}D_{3}$ & $M$ & $2.840$ & $3.075$ & $3.225$ & $3.310$ & $3.344$ & $3.351$ & $3.354$ & $3.355$ & $3.356$ & $3.357$ \\
 & $r$ & $1.00$ & $1.62$ & $2.47$ & $3.93$ & $7.73$ & $14.82$ & $23.30$ & $33.53$ & $46.96$ & $60.38$ \\
\end{tabular}
\end{ruledtabular}
\end{table*}

\renewcommand{\tabcolsep}{0.11cm}
\renewcommand{\arraystretch}{1.05}
\begin{table*}[t]
\caption{Extended $B_s$ spectrum for the $nS$, $nP$, and $nD$ families with $n=1$--$10$. For each channel the calculated mass $M$ (GeV) and RMS radius $r$ (fm) are listed.}
\label{tab:Bs-spectra}
\begin{ruledtabular}
\begin{tabular}{cl*{10}{c}}
$^{2S+1}L_{J}$ & & \multicolumn{10}{c}{$n$} \\
 & & $1$ & $2$ & $3$ & $4$ & $5$ & $6$ & $7$ & $8$ & $9$ & $10$ \\ \hline
$^{1}S_{0}$ & $M$ & $5.370$ & $5.927$ & $6.252$ & $6.466$ & $6.609$ & $6.701$ & $6.751$ & $6.770$ & $6.776$ & $6.778$ \\
 & $r$ & $0.36$ & $0.76$ & $1.21$ & $1.74$ & $2.43$ & $3.45$ & $5.34$ & $9.52$ & $15.68$ & $23.13$ \\
$^{3}S_{1}$ & $M$ & $5.424$ & $5.951$ & $6.266$ & $6.475$ & $6.615$ & $6.704$ & $6.753$ & $6.770$ & $6.776$ & $6.778$ \\
 & $r$ & $0.38$ & $0.79$ & $1.23$ & $1.77$ & $2.46$ & $3.51$ & $5.47$ & $9.79$ & $16.00$ & $23.50$ \\
$^{1}P_{1}$ & $M$ & $5.820$ & $6.171$ & $6.407$ & $6.569$ & $6.676$ & $6.739$ & $6.767$ & $6.775$ & $6.778$ & $6.780$ \\
 & $r$ & $0.59$ & $1.02$ & $1.51$ & $2.12$ & $3.00$ & $4.49$ & $7.82$ & $13.71$ & $20.86$ & $29.49$ \\
$^{3}P_{0}$ & $M$ & $5.782$ & $6.152$ & $6.395$ & $6.561$ & $6.671$ & $6.737$ & $6.766$ & $6.774$ & $6.778$ & $6.780$ \\
 & $r$ & $0.54$ & $0.98$ & $1.47$ & $2.08$ & $2.93$ & $4.37$ & $7.52$ & $13.31$ & $20.37$ & $28.86$ \\
$^{3}P_{1}$ & $M$ & $5.816$ & $6.169$ & $6.406$ & $6.568$ & $6.675$ & $6.739$ & $6.767$ & $6.775$ & $6.778$ & $6.780$ \\
 & $r$ & $0.58$ & $1.01$ & $1.50$ & $2.12$ & $2.99$ & $4.47$ & $7.77$ & $13.65$ & $20.79$ & $29.40$ \\
$^{3}P_{2}$ & $M$ & $5.840$ & $6.185$ & $6.417$ & $6.576$ & $6.680$ & $6.742$ & $6.768$ & $6.775$ & $6.778$ & $6.780$ \\
 & $r$ & $0.62$ & $1.05$ & $1.54$ & $2.17$ & $3.07$ & $4.61$ & $8.11$ & $14.10$ & $21.33$ & $30.09$ \\
$^{1}D_{2}$ & $M$ & $6.107$ & $6.358$ & $6.534$ & $6.654$ & $6.729$ & $6.764$ & $6.774$ & $6.778$ & $6.780$ & $6.781$ \\
 & $r$ & $0.83$ & $1.30$ & $1.87$ & $2.64$ & $3.89$ & $6.63$ & $12.51$ & $19.72$ & $28.23$ & $38.07$ \\
$^{3}D_{1}$ & $M$ & $6.105$ & $6.355$ & $6.531$ & $6.651$ & $6.727$ & $6.764$ & $6.774$ & $6.778$ & $6.779$ & $6.781$ \\
 & $r$ & $0.80$ & $1.27$ & $1.84$ & $2.60$ & $3.82$ & $6.46$ & $12.24$ & $19.39$ & $27.85$ & $37.69$ \\
$^{3}D_{2}$ & $M$ & $6.110$ & $6.360$ & $6.535$ & $6.654$ & $6.729$ & $6.764$ & $6.774$ & $6.778$ & $6.780$ & $6.781$ \\
 & $r$ & $0.82$ & $1.30$ & $1.87$ & $2.64$ & $3.89$ & $6.62$ & $12.50$ & $19.70$ & $28.21$ & $38.05$ \\
$^{3}D_{3}$ & $M$ & $6.107$ & $6.360$ & $6.537$ & $6.656$ & $6.730$ & $6.765$ & $6.775$ & $6.778$ & $6.780$ & $6.781$ \\
 & $r$ & $0.85$ & $1.32$ & $1.89$ & $2.68$ & $3.95$ & $6.76$ & $12.72$ & $19.96$ & $28.52$ & $38.36$ \\
\end{tabular}
\end{ruledtabular}
\end{table*}

\renewcommand{\tabcolsep}{0.11cm}
\renewcommand{\arraystretch}{1.05}
\begin{table*}[t]
\caption{Extended $B_c$ spectrum for the $nS$, $nP$, and $nD$ families with $n=1$--$14$. For each channel the calculated mass $M$ (GeV) and RMS radius $r$ (fm) are listed.}
\label{tab:Bc-spectra}
\begin{ruledtabular}
\begin{tabular}{cl*{14}{c}}
$^{2S+1}L_{J}$ & & \multicolumn{14}{c}{$n$} \\
 & & $1$ & $2$ & $3$ & $4$ & $5$ & $6$ & $7$ & $8$ & $9$ & $10$ & $11$ & $12$ & $13$ & $14$ \\ \hline
$^{1}S_{0}$ & $M$ & $6.277$ & $6.870$ & $7.221$ & $7.469$ & $7.657$ & $7.804$ & $7.932$ & $8.032$ & $8.087$ & $8.131$ & $8.161$ & $8.174$ & $8.179$ & $8.181$ \\
 & $r$ & $0.25$ & $0.56$ & $0.87$ & $1.19$ & $1.54$ & $2.01$ & $2.76$ & $3.33$ & $3.72$ & $5.46$ & $7.65$ & $10.42$ & $14.27$ & $18.87$ \\
$^{3}S_{1}$ & $M$ & $6.341$ & $6.899$ & $7.240$ & $7.483$ & $7.667$ & $7.811$ & $7.937$ & $8.039$ & $8.098$ & $8.134$ & $8.162$ & $8.174$ & $8.179$ & $8.181$ \\
 & $r$ & $0.27$ & $0.58$ & $0.89$ & $1.21$ & $1.55$ & $2.03$ & $2.81$ & $3.63$ & $4.49$ & $5.38$ & $7.60$ & $10.43$ & $14.29$ & $18.89$ \\
$^{1}P_{1}$ & $M$ & $6.759$ & $7.135$ & $7.401$ & $7.603$ & $7.760$ & $7.881$ & $7.985$ & $8.076$ & $8.136$ & $8.165$ & $8.176$ & $8.180$ & $8.182$ & $8.183$ \\
 & $r$ & $0.44$ & $0.75$ & $1.07$ & $1.41$ & $1.77$ & $2.19$ & $3.11$ & $4.41$ & $6.04$ & $8.13$ & $11.14$ & $15.22$ & $20.05$ & $26.42$ \\
$^{3}P_{0}$ & $M$ & $6.701$ & $7.100$ & $7.376$ & $7.584$ & $7.745$ & $7.870$ & $7.981$ & $8.074$ & $8.134$ & $8.164$ & $8.175$ & $8.180$ & $8.182$ & $8.183$ \\
 & $r$ & $0.41$ & $0.72$ & $1.04$ & $1.37$ & $1.72$ & $2.19$ & $3.12$ & $4.36$ & $5.93$ & $7.97$ & $10.94$ & $15.02$ & $19.91$ & $26.42$ \\
$^{3}P_{1}$ & $M$ & $6.748$ & $7.129$ & $7.397$ & $7.599$ & $7.757$ & $7.879$ & $7.985$ & $8.076$ & $8.136$ & $8.165$ & $8.176$ & $8.180$ & $8.182$ & $8.183$ \\
 & $r$ & $0.44$ & $0.75$ & $1.06$ & $1.40$ & $1.76$ & $2.19$ & $3.12$ & $4.40$ & $6.03$ & $8.12$ & $11.12$ & $15.20$ & $20.04$ & $26.42$ \\
$^{3}P_{2}$ & $M$ & $6.784$ & $7.152$ & $7.414$ & $7.613$ & $7.769$ & $7.887$ & $7.988$ & $8.077$ & $8.136$ & $8.165$ & $8.176$ & $8.180$ & $8.182$ & $8.183$ \\
 & $r$ & $0.46$ & $0.77$ & $1.09$ & $1.44$ & $1.80$ & $2.18$ & $3.10$ & $4.42$ & $6.06$ & $8.15$ & $11.17$ & $15.25$ & $20.07$ & $26.42$ \\
$^{1}D_{2}$ & $M$ & $7.045$ & $7.331$ & $7.546$ & $7.714$ & $7.852$ & $7.957$ & $8.031$ & $8.101$ & $8.148$ & $8.170$ & $8.178$ & $8.181$ & $8.183$ & $8.184$ \\
 & $r$ & $0.62$ & $0.94$ & $1.26$ & $1.64$ & $2.13$ & $2.72$ & $3.42$ & $4.66$ & $6.43$ & $8.72$ & $12.19$ & $16.53$ & $21.48$ & $27.91$ \\
$^{3}D_{1}$ & $M$ & $7.025$ & $7.316$ & $7.534$ & $7.705$ & $7.843$ & $7.947$ & $8.026$ & $8.100$ & $8.148$ & $8.170$ & $8.178$ & $8.181$ & $8.183$ & $8.184$ \\
 & $r$ & $0.59$ & $0.91$ & $1.24$ & $1.61$ & $2.06$ & $2.58$ & $3.22$ & $4.68$ & $6.43$ & $8.72$ & $12.18$ & $16.52$ & $21.47$ & $27.91$ \\
$^{3}D_{2}$ & $M$ & $7.042$ & $7.329$ & $7.544$ & $7.713$ & $7.851$ & $7.956$ & $8.030$ & $8.101$ & $8.148$ & $8.170$ & $8.178$ & $8.181$ & $8.183$ & $8.184$ \\
 & $r$ & $0.61$ & $0.93$ & $1.26$ & $1.64$ & $2.12$ & $2.70$ & $3.15$ & $4.67$ & $6.43$ & $8.72$ & $12.19$ & $16.53$ & $21.48$ & $27.91$ \\
$^{3}D_{3}$ & $M$ & $7.057$ & $7.340$ & $7.554$ & $7.720$ & $7.857$ & $7.964$ & $8.035$ & $8.102$ & $8.148$ & $8.170$ & $8.178$ & $8.181$ & $8.183$ & $8.184$ \\
 & $r$ & $0.63$ & $0.95$ & $1.28$ & $1.66$ & $2.17$ & $2.81$ & $3.57$ & $4.64$ & $6.42$ & $8.72$ & $12.20$ & $16.54$ & $21.48$ & $27.90$ \\
\end{tabular}
\end{ruledtabular}
\end{table*}

\renewcommand{\tabcolsep}{0.12cm}
\renewcommand{\arraystretch}{1.1}
\begin{table*}[!htbp]
\caption{Numerical summary of the termination boundaries for the extended spectra. The safe range ends at the state immediately before the first caution state. The column $\Delta M_{\textrm{caution}}$ gives the mass gap in MeV, and $\Delta r_{\textrm{caution}}$ gives the RMS-radius jump in fm at the caution onset. The next two columns list the changepoints inferred from the CUSUM and two-segment analyses of the indicator $f_{n}$. The column $n_{\chi<1}$ gives the first index at which $\chi_{n}<1$, and $\chi_{\chi<1}$ is the corresponding value of $\chi_{n}$.}
\label{tab:termination}
\begin{ruledtabular}
\begin{tabular}{llcccccccc}
 & & \multicolumn{4}{c}{Threshold-based} & \multicolumn{2}{c}{Changepoint-based} & \multicolumn{2}{c}{$\chi$-based} \\
\cline{3-6}\cline{7-8}\cline{9-10}
\multicolumn{2}{l}{Channel} & $n_{\textrm{safe}}$ & $n_{\textrm{caution}}$ & $\Delta M_{\textrm{caution}}$ & $\Delta r_{\textrm{caution}}$
 & $n_{\textrm{CUSUM}}$ & $n_{\textrm{2-seg}}$ & $n_{\chi<1}$ & $\chi_{\chi<1}$ \\ \hline
$D_s$ & $^{1}S_{0}$ & $6$ & $7$ & $5.60$ & $6.82$ & $6$ & $7$ & $8$ & $0.474$ \\
 & $^{3}S_{1}$ & $6$ & $7$ & $5.17$ & $6.94$ & $6$ & $7$ & $7$ & $0.946$ \\
 & $^{1}P_{1}$ & $5$ & $6$ & $7.17$ & $6.63$ & $6$ & $6$ & $7$ & $0.540$ \\
 & $^{3}P_{0}$ & $5$ & $6$ & $9.48$ & $6.26$ & $6$ & $7$ & $7$ & $0.641$ \\
 & $^{3}P_{1}$ & $5$ & $6$ & $7.59$ & $6.56$ & $6$ & $7$ & $7$ & $0.559$ \\
 & $^{3}P_{2}$ & $5$ & $6$ & $6.15$ & $6.83$ & $6$ & $6$ & $7$ & $0.493$ \\
 & $^{1}D_{2}$ & $4$ & $5$ & $8.36$ & $6.89$ & $5$ & $6$ & $6$ & $0.592$ \\
 & $^{3}D_{1}$ & $5$ & $6$ & $3.43$ & $8.00$ & $5$ & $6$ & $6$ & $0.723$ \\
 & $^{3}D_{2}$ & $4$ & $5$ & $8.78$ & $6.83$ & $5$ & $6$ & $6$ & $0.611$ \\
 & $^{3}D_{3}$ & $4$ & $5$ & $7.05$ & $7.10$ & $5$ & $6$ & $6$ & $0.530$ \\
$B_s$ & $^{1}S_{0}$ & $7$ & $8$ & $5.68$ & $6.16$ & $7$ & $8$ & $8$ & $0.912$ \\
 & $^{3}S_{1}$ & $7$ & $8$ & $5.42$ & $6.21$ & $7$ & $8$ & $8$ & $0.883$ \\
 & $^{1}P_{1}$ & $6$ & $7$ & $7.76$ & $5.90$ & $7$ & $8$ & $8$ & $0.539$ \\
 & $^{3}P_{0}$ & $6$ & $7$ & $8.38$ & $5.79$ & $7$ & $8$ & $8$ & $0.565$ \\
 & $^{3}P_{1}$ & $6$ & $7$ & $7.84$ & $5.88$ & $7$ & $8$ & $8$ & $0.543$ \\
 & $^{3}P_{2}$ & $6$ & $7$ & $7.23$ & $5.99$ & $7$ & $7$ & $8$ & $0.517$ \\
 & $^{1}D_{2}$ & $5$ & $6$ & $9.97$ & $5.89$ & $6$ & $7$ & $7$ & $0.632$ \\
 & $^{3}D_{1}$ & $6$ & $7$ & $3.45$ & $7.15$ & $6$ & $7$ & $7$ & $0.654$ \\
 & $^{3}D_{2}$ & $5$ & $6$ & $9.99$ & $5.88$ & $6$ & $7$ & $7$ & $0.632$ \\
 & $^{3}D_{3}$ & $5$ & $6$ & $9.55$ & $5.96$ & $6$ & $7$ & $7$ & $0.616$ \\
$B_c$ & $^{1}S_{0}$ & $11$ & $12$ & $4.77$ & $3.85$ & $9$ & $10$ & $12$ & $0.676$ \\
 & $^{3}S_{1}$ & $11$ & $12$ & $4.70$ & $3.86$ & $9$ & $10$ & $12$ & $0.657$ \\
 & $^{1}P_{1}$ & $10$ & $11$ & $4.03$ & $4.08$ & $9$ & $10$ & $11$ & $0.606$ \\
 & $^{3}P_{0}$ & $10$ & $11$ & $4.21$ & $4.08$ & $9$ & $10$ & $11$ & $0.625$ \\
 & $^{3}P_{1}$ & $10$ & $11$ & $4.04$ & $4.08$ & $9$ & $10$ & $11$ & $0.608$ \\
 & $^{3}P_{2}$ & $10$ & $11$ & $3.99$ & $4.09$ & $9$ & $10$ & $11$ & $0.599$ \\
 & $^{1}D_{2}$ & $9$ & $10$ & $7.61$ & $3.47$ & $9$ & $10$ & $10$ & $0.962$ \\
 & $^{3}D_{1}$ & $9$ & $10$ & $7.65$ & $3.46$ & $9$ & $10$ & $10$ & $0.969$ \\
 & $^{3}D_{2}$ & $9$ & $10$ & $7.62$ & $3.47$ & $9$ & $10$ & $10$ & $0.963$ \\
 & $^{3}D_{3}$ & $9$ & $10$ & $7.58$ & $3.47$ & $9$ & $10$ & $10$ & $0.957$ \\
\end{tabular}
\end{ruledtabular}
\end{table*}

\begin{acknowledgments}
This work is supported by the National Natural Science Foundation of China (Grant No.~12247101), 
the Fundamental Research Funds for the Central Universities (Grant No.~lzujbky-2025-jdzx07), 
the Natural Science Foundation of Gansu Province (Grant No.~25JRRA799), 
and the ``111 Center'' under Grant No.~B20063.

AI Usage Declaration: The initial version of the manuscript was drafted with the assistance of Grok-4.6 \cite{xaigrok} and subsequently revised by the authors. Data processing and visualization were carried out with Grok-4.6. The authors reviewed and edited the output as needed and take full responsibility for the content of the published article.

Data Availability: The codebase and numerical tests are available at GitHub \cite{gemstore}.
\end{acknowledgments}


\begin{thebibliography}{60}%
\makeatletter
\providecommand \@ifxundefined [1]{%
 \@ifx{#1\undefined}
}%
\providecommand \@ifnum [1]{%
 \ifnum #1\expandafter \@firstoftwo
 \else \expandafter \@secondoftwo
 \fi
}%
\providecommand \@ifx [1]{%
 \ifx #1\expandafter \@firstoftwo
 \else \expandafter \@secondoftwo
 \fi
}%
\providecommand \natexlab [1]{#1}%
\providecommand \enquote  [1]{``#1''}%
\providecommand \bibnamefont  [1]{#1}%
\providecommand \bibfnamefont [1]{#1}%
\providecommand \citenamefont [1]{#1}%
\providecommand \href@noop [0]{\@secondoftwo}%
\providecommand \href [0]{\begingroup \@sanitize@url \@href}%
\providecommand \@href[1]{\@@startlink{#1}\@@href}%
\providecommand \@@href[1]{\endgroup#1\@@endlink}%
\providecommand \@sanitize@url [0]{\catcode `\\12\catcode `\$12\catcode `\&12\catcode `\#12\catcode `\^12\catcode `\_12\catcode `\%12\relax}%
\providecommand \@@startlink[1]{}%
\providecommand \@@endlink[0]{}%
\providecommand \url  [0]{\begingroup\@sanitize@url \@url }%
\providecommand \@url [1]{\endgroup\@href {#1}{\urlprefix }}%
\providecommand \urlprefix  [0]{URL }%
\providecommand \Eprint [0]{\href }%
\providecommand \doibase [0]{http://dx.doi.org/}%
\providecommand \selectlanguage [0]{\@gobble}%
\providecommand \bibinfo  [0]{\@secondoftwo}%
\providecommand \bibfield  [0]{\@secondoftwo}%
\providecommand \translation [1]{[#1]}%
\providecommand \BibitemOpen [0]{}%
\providecommand \bibitemStop [0]{}%
\providecommand \bibitemNoStop [0]{.\EOS\space}%
\providecommand \EOS [0]{\spacefactor3000\relax}%
\providecommand \BibitemShut  [1]{\csname bibitem#1\endcsname}%
\let\auto@bib@innerbib\@empty
\bibitem [{\citenamefont {Eichten}\ \emph {et~al.}(1975)\citenamefont {Eichten}, \citenamefont {Gottfried}, \citenamefont {Kinoshita}, \citenamefont {Kogut}, \citenamefont {Lane},\ and\ \citenamefont {Yan}}]{Eichten:1974af}%
  \BibitemOpen
  \bibfield  {author} {\bibinfo {author} {\bibfnamefont {E.}~\bibnamefont {Eichten}}, \bibinfo {author} {\bibfnamefont {K.}~\bibnamefont {Gottfried}}, \bibinfo {author} {\bibfnamefont {T.}~\bibnamefont {Kinoshita}}, \bibinfo {author} {\bibfnamefont {J.~B.}\ \bibnamefont {Kogut}}, \bibinfo {author} {\bibfnamefont {K.~D.}\ \bibnamefont {Lane}}, \ and\ \bibinfo {author} {\bibfnamefont {T.-M.}\ \bibnamefont {Yan}},\ }\href {\doibase 10.1103/PhysRevLett.34.369} {\bibfield  {journal} {\bibinfo  {journal} {Phys. Rev. Lett.}\ }\textbf {\bibinfo {volume} {34}},\ \bibinfo {pages} {369} (\bibinfo {year} {1975})},\ \bibinfo {note} {[Erratum: Phys.Rev.Lett. 36, 1276 (1976)]}\BibitemShut {NoStop}%
\bibitem [{\citenamefont {Eichten}\ \emph {et~al.}(1978)\citenamefont {Eichten}, \citenamefont {Gottfried}, \citenamefont {Kinoshita}, \citenamefont {Lane},\ and\ \citenamefont {Yan}}]{Eichten:1978tg}%
  \BibitemOpen
  \bibfield  {author} {\bibinfo {author} {\bibfnamefont {E.}~\bibnamefont {Eichten}}, \bibinfo {author} {\bibfnamefont {K.}~\bibnamefont {Gottfried}}, \bibinfo {author} {\bibfnamefont {T.}~\bibnamefont {Kinoshita}}, \bibinfo {author} {\bibfnamefont {K.~D.}\ \bibnamefont {Lane}}, \ and\ \bibinfo {author} {\bibfnamefont {T.-M.}\ \bibnamefont {Yan}},\ }\href {\doibase 10.1103/PhysRevD.17.3090} {\bibfield  {journal} {\bibinfo  {journal} {Phys. Rev. D}\ }\textbf {\bibinfo {volume} {17}},\ \bibinfo {pages} {3090} (\bibinfo {year} {1978})},\ \bibinfo {note} {[Erratum: Phys.Rev.D 21, 313 (1980)]}\BibitemShut {NoStop}%
\bibitem [{\citenamefont {Eichten}\ \emph {et~al.}(1980)\citenamefont {Eichten}, \citenamefont {Gottfried}, \citenamefont {Kinoshita}, \citenamefont {Lane},\ and\ \citenamefont {Yan}}]{Eichten:1979ms}%
  \BibitemOpen
  \bibfield  {author} {\bibinfo {author} {\bibfnamefont {E.}~\bibnamefont {Eichten}}, \bibinfo {author} {\bibfnamefont {K.}~\bibnamefont {Gottfried}}, \bibinfo {author} {\bibfnamefont {T.}~\bibnamefont {Kinoshita}}, \bibinfo {author} {\bibfnamefont {K.~D.}\ \bibnamefont {Lane}}, \ and\ \bibinfo {author} {\bibfnamefont {T.-M.}\ \bibnamefont {Yan}},\ }\href {\doibase 10.1103/PhysRevD.21.203} {\bibfield  {journal} {\bibinfo  {journal} {Phys. Rev. D}\ }\textbf {\bibinfo {volume} {21}},\ \bibinfo {pages} {203} (\bibinfo {year} {1980})}\BibitemShut {NoStop}%
\bibitem [{\citenamefont {Bali}\ \emph {et~al.}(2005)\citenamefont {Bali}, \citenamefont {Neff}, \citenamefont {Duessel}, \citenamefont {Lippert},\ and\ \citenamefont {Schilling}}]{Bali:2005fu}%
  \BibitemOpen
  \bibfield  {author} {\bibinfo {author} {\bibfnamefont {G.~S.}\ \bibnamefont {Bali}}, \bibinfo {author} {\bibfnamefont {H.}~\bibnamefont {Neff}}, \bibinfo {author} {\bibfnamefont {T.}~\bibnamefont {Duessel}}, \bibinfo {author} {\bibfnamefont {T.}~\bibnamefont {Lippert}}, \ and\ \bibinfo {author} {\bibfnamefont {K.}~\bibnamefont {Schilling}} (\bibinfo {collaboration} {SESAM}),\ }\href {\doibase 10.1103/PhysRevD.71.114513} {\bibfield  {journal} {\bibinfo  {journal} {Phys. Rev. D}\ }\textbf {\bibinfo {volume} {71}},\ \bibinfo {pages} {114513} (\bibinfo {year} {2005})},\ \Eprint {http://arxiv.org/abs/hep-lat/0505012} {arXiv:hep-lat/0505012} \BibitemShut {NoStop}%
\bibitem [{\citenamefont {Castorina}\ \emph {et~al.}(2007)\citenamefont {Castorina}, \citenamefont {Kharzeev},\ and\ \citenamefont {Satz}}]{Castorina:2007eb}%
  \BibitemOpen
  \bibfield  {author} {\bibinfo {author} {\bibfnamefont {P.}~\bibnamefont {Castorina}}, \bibinfo {author} {\bibfnamefont {D.}~\bibnamefont {Kharzeev}}, \ and\ \bibinfo {author} {\bibfnamefont {H.}~\bibnamefont {Satz}},\ }\href {\doibase 10.1140/epjc/s10052-007-0368-6} {\bibfield  {journal} {\bibinfo  {journal} {Eur. Phys. J. C}\ }\textbf {\bibinfo {volume} {52}},\ \bibinfo {pages} {187} (\bibinfo {year} {2007})},\ \Eprint {http://arxiv.org/abs/0704.1426} {arXiv:0704.1426 [hep-ph]} \BibitemShut {NoStop}%
\bibitem [{\citenamefont {Bulava}\ \emph {et~al.}(2019)\citenamefont {Bulava}, \citenamefont {H\"orz}, \citenamefont {Knechtli}, \citenamefont {Koch}, \citenamefont {Moir}, \citenamefont {Morningstar},\ and\ \citenamefont {Peardon}}]{Bulava:2019iut}%
  \BibitemOpen
  \bibfield  {author} {\bibinfo {author} {\bibfnamefont {J.}~\bibnamefont {Bulava}}, \bibinfo {author} {\bibfnamefont {B.}~\bibnamefont {H\"orz}}, \bibinfo {author} {\bibfnamefont {F.}~\bibnamefont {Knechtli}}, \bibinfo {author} {\bibfnamefont {V.}~\bibnamefont {Koch}}, \bibinfo {author} {\bibfnamefont {G.}~\bibnamefont {Moir}}, \bibinfo {author} {\bibfnamefont {C.}~\bibnamefont {Morningstar}}, \ and\ \bibinfo {author} {\bibfnamefont {M.}~\bibnamefont {Peardon}},\ }\href {\doibase 10.1016/j.physletb.2019.05.018} {\bibfield  {journal} {\bibinfo  {journal} {Phys. Lett. B}\ }\textbf {\bibinfo {volume} {793}},\ \bibinfo {pages} {493} (\bibinfo {year} {2019})},\ \Eprint {http://arxiv.org/abs/1902.04006} {arXiv:1902.04006 [hep-lat]} \BibitemShut {NoStop}%
\bibitem [{\citenamefont {Jiang}\ \emph {et~al.}(2023)\citenamefont {Jiang}, \citenamefont {Chen}, \citenamefont {Qin},\ and\ \citenamefont {Martin~Contreras}}]{Jiang:2023lmj}%
  \BibitemOpen
  \bibfield  {author} {\bibinfo {author} {\bibfnamefont {J.-J.}\ \bibnamefont {Jiang}}, \bibinfo {author} {\bibfnamefont {X.}~\bibnamefont {Chen}}, \bibinfo {author} {\bibfnamefont {J.}~\bibnamefont {Qin}}, \ and\ \bibinfo {author} {\bibfnamefont {M.~A.}\ \bibnamefont {Martin~Contreras}},\ }\href {\doibase 10.1103/PhysRevD.108.126002} {\bibfield  {journal} {\bibinfo  {journal} {Phys. Rev. D}\ }\textbf {\bibinfo {volume} {108}},\ \bibinfo {pages} {126002} (\bibinfo {year} {2023})},\ \Eprint {http://arxiv.org/abs/2310.04983} {arXiv:2310.04983 [hep-ph]} \BibitemShut {NoStop}%
\bibitem [{\citenamefont {Kou}\ and\ \citenamefont {Chen}(2024)}]{Kou:2024dml}%
  \BibitemOpen
  \bibfield  {author} {\bibinfo {author} {\bibfnamefont {W.}~\bibnamefont {Kou}}\ and\ \bibinfo {author} {\bibfnamefont {X.}~\bibnamefont {Chen}},\ }\href {\doibase 10.1016/j.physletb.2024.138942} {\bibfield  {journal} {\bibinfo  {journal} {Phys. Lett. B}\ }\textbf {\bibinfo {volume} {856}},\ \bibinfo {pages} {138942} (\bibinfo {year} {2024})},\ \Eprint {http://arxiv.org/abs/2405.18697} {arXiv:2405.18697 [hep-ph]} \BibitemShut {NoStop}%
\bibitem [{\citenamefont {Chen}\ \emph {et~al.}(2017)\citenamefont {Chen}, \citenamefont {Chen}, \citenamefont {Liu}, \citenamefont {Liu},\ and\ \citenamefont {Zhu}}]{Chen:2017rpp}%
  \BibitemOpen
  \bibfield  {author} {\bibinfo {author} {\bibfnamefont {H.-X.}\ \bibnamefont {Chen}}, \bibinfo {author} {\bibfnamefont {W.}~\bibnamefont {Chen}}, \bibinfo {author} {\bibfnamefont {X.}~\bibnamefont {Liu}}, \bibinfo {author} {\bibfnamefont {Y.-R.}\ \bibnamefont {Liu}}, \ and\ \bibinfo {author} {\bibfnamefont {S.-L.}\ \bibnamefont {Zhu}},\ }\href {\doibase 10.1088/1361-6633/aa6420} {\bibfield  {journal} {\bibinfo  {journal} {Rept. Prog. Phys.}\ }\textbf {\bibinfo {volume} {80}},\ \bibinfo {pages} {076201} (\bibinfo {year} {2017})},\ \Eprint {http://arxiv.org/abs/1609.08928} {arXiv:1609.08928 [hep-ph]} \BibitemShut {NoStop}%
\bibitem [{\citenamefont {Navas}\ \emph {et~al.}(2024)\citenamefont {Navas} \emph {et~al.}}]{ParticleDataGroup:2024cfk}%
  \BibitemOpen
  \bibfield  {author} {\bibinfo {author} {\bibfnamefont {S.}~\bibnamefont {Navas}} \emph {et~al.} (\bibinfo {collaboration} {Particle Data Group}),\ }\href {\doibase 10.1103/PhysRevD.110.030001} {\bibfield  {journal} {\bibinfo  {journal} {Phys. Rev. D}\ }\textbf {\bibinfo {volume} {110}},\ \bibinfo {pages} {030001} (\bibinfo {year} {2024})}\BibitemShut {NoStop}%
\bibitem [{\citenamefont {Aubert}\ \emph {et~al.}(2003)\citenamefont {Aubert} \emph {et~al.}}]{BaBar:2003ds2317}%
  \BibitemOpen
  \bibfield  {author} {\bibinfo {author} {\bibfnamefont {B.}~\bibnamefont {Aubert}} \emph {et~al.} (\bibinfo {collaboration} {BaBar}),\ }\href {\doibase 10.1103/PhysRevLett.90.242001} {\bibfield  {journal} {\bibinfo  {journal} {Phys. Rev. Lett.}\ }\textbf {\bibinfo {volume} {90}},\ \bibinfo {pages} {242001} (\bibinfo {year} {2003})},\ \Eprint {http://arxiv.org/abs/hep-ex/0304021} {arXiv:hep-ex/0304021} \BibitemShut {NoStop}%
\bibitem [{\citenamefont {Besson}\ \emph {et~al.}(2003)\citenamefont {Besson} \emph {et~al.}}]{CLEO:2003ds2460}%
  \BibitemOpen
  \bibfield  {author} {\bibinfo {author} {\bibfnamefont {D.}~\bibnamefont {Besson}} \emph {et~al.} (\bibinfo {collaboration} {CLEO}),\ }\href {\doibase 10.1103/PhysRevD.68.032002} {\bibfield  {journal} {\bibinfo  {journal} {Phys. Rev. D}\ }\textbf {\bibinfo {volume} {68}},\ \bibinfo {pages} {032002} (\bibinfo {year} {2003})},\ \Eprint {http://arxiv.org/abs/hep-ex/0305100} {arXiv:hep-ex/0305100} \BibitemShut {NoStop}%
\bibitem [{\citenamefont {Krokovny}\ \emph {et~al.}(2003)\citenamefont {Krokovny} \emph {et~al.}}]{Belle:2003ds2457}%
  \BibitemOpen
  \bibfield  {author} {\bibinfo {author} {\bibfnamefont {P.}~\bibnamefont {Krokovny}} \emph {et~al.} (\bibinfo {collaboration} {Belle}),\ }\href {\doibase 10.1103/PhysRevLett.91.262002} {\bibfield  {journal} {\bibinfo  {journal} {Phys. Rev. Lett.}\ }\textbf {\bibinfo {volume} {91}},\ \bibinfo {pages} {262002} (\bibinfo {year} {2003})},\ \Eprint {http://arxiv.org/abs/hep-ex/0308019} {arXiv:hep-ex/0308019} \BibitemShut {NoStop}%
\bibitem [{\citenamefont {Albrecht}\ \emph {et~al.}(1989)\citenamefont {Albrecht} \emph {et~al.}}]{ARGUS:1989ds1}%
  \BibitemOpen
  \bibfield  {author} {\bibinfo {author} {\bibfnamefont {H.}~\bibnamefont {Albrecht}} \emph {et~al.} (\bibinfo {collaboration} {ARGUS}),\ }\href {\doibase 10.1016/0370-2693(89)91672-9} {\bibfield  {journal} {\bibinfo  {journal} {Phys. Lett. B}\ }\textbf {\bibinfo {volume} {230}},\ \bibinfo {pages} {162} (\bibinfo {year} {1989})}\BibitemShut {NoStop}%
\bibitem [{\citenamefont {Kubota}\ \emph {et~al.}(1994)\citenamefont {Kubota} \emph {et~al.}}]{CLEO:1994ds2}%
  \BibitemOpen
  \bibfield  {author} {\bibinfo {author} {\bibfnamefont {Y.}~\bibnamefont {Kubota}} \emph {et~al.} (\bibinfo {collaboration} {CLEO}),\ }\href {\doibase 10.1103/PhysRevLett.72.1972} {\bibfield  {journal} {\bibinfo  {journal} {Phys. Rev. Lett.}\ }\textbf {\bibinfo {volume} {72}},\ \bibinfo {pages} {1972} (\bibinfo {year} {1994})},\ \Eprint {http://arxiv.org/abs/hep-ph/9403325} {arXiv:hep-ph/9403325} \BibitemShut {NoStop}%
\bibitem [{\citenamefont {Brodzicka}\ \emph {et~al.}(2008)\citenamefont {Brodzicka} \emph {et~al.}}]{Belle:2008ds2700}%
  \BibitemOpen
  \bibfield  {author} {\bibinfo {author} {\bibfnamefont {J.}~\bibnamefont {Brodzicka}} \emph {et~al.} (\bibinfo {collaboration} {Belle}),\ }\href {\doibase 10.1103/PhysRevLett.100.092001} {\bibfield  {journal} {\bibinfo  {journal} {Phys. Rev. Lett.}\ }\textbf {\bibinfo {volume} {100}},\ \bibinfo {pages} {092001} (\bibinfo {year} {2008})},\ \Eprint {http://arxiv.org/abs/0707.3491} {arXiv:0707.3491 [hep-ex]} \BibitemShut {NoStop}%
\bibitem [{\citenamefont {Aaij}\ \emph {et~al.}(2014)\citenamefont {Aaij} \emph {et~al.}}]{LHCb:2014ds2860}%
  \BibitemOpen
  \bibfield  {author} {\bibinfo {author} {\bibfnamefont {R.}~\bibnamefont {Aaij}} \emph {et~al.} (\bibinfo {collaboration} {LHCb}),\ }\href {\doibase 10.1103/PhysRevLett.113.162001} {\bibfield  {journal} {\bibinfo  {journal} {Phys. Rev. Lett.}\ }\textbf {\bibinfo {volume} {113}},\ \bibinfo {pages} {162001} (\bibinfo {year} {2014})},\ \Eprint {http://arxiv.org/abs/1407.7574} {arXiv:1407.7574 [hep-ex]} \BibitemShut {NoStop}%
\bibitem [{\citenamefont {Aaij}\ \emph {et~al.}(2021{\natexlab{a}})\citenamefont {Aaij} \emph {et~al.}}]{LHCb:2020gey}%
  \BibitemOpen
  \bibfield  {author} {\bibinfo {author} {\bibfnamefont {R.}~\bibnamefont {Aaij}} \emph {et~al.} (\bibinfo {collaboration} {LHCb}),\ }\href {\doibase 10.1103/PhysRevLett.126.122002} {\bibfield  {journal} {\bibinfo  {journal} {Phys. Rev. Lett.}\ }\textbf {\bibinfo {volume} {126}},\ \bibinfo {pages} {122002} (\bibinfo {year} {2021}{\natexlab{a}})},\ \Eprint {http://arxiv.org/abs/2011.09112} {arXiv:2011.09112 [hep-ex]} \BibitemShut {NoStop}%
\bibitem [{\citenamefont {Aaij}\ \emph {et~al.}(2026)\citenamefont {Aaij} \emph {et~al.}}]{LHCb:2026ds2933}%
  \BibitemOpen
  \bibfield  {author} {\bibinfo {author} {\bibfnamefont {R.}~\bibnamefont {Aaij}} \emph {et~al.} (\bibinfo {collaboration} {LHCb}),\ }\href@noop {} {\  (\bibinfo {year} {2026})},\ \Eprint {http://arxiv.org/abs/2604.21257} {arXiv:2604.21257 [hep-ex]} \BibitemShut {NoStop}%
\bibitem [{\citenamefont {Aaltonen}\ \emph {et~al.}(2008)\citenamefont {Aaltonen} \emph {et~al.}}]{CDF:2008bs}%
  \BibitemOpen
  \bibfield  {author} {\bibinfo {author} {\bibfnamefont {T.}~\bibnamefont {Aaltonen}} \emph {et~al.} (\bibinfo {collaboration} {CDF}),\ }\href {\doibase 10.1103/PhysRevLett.100.082001} {\bibfield  {journal} {\bibinfo  {journal} {Phys. Rev. Lett.}\ }\textbf {\bibinfo {volume} {100}},\ \bibinfo {pages} {082001} (\bibinfo {year} {2008})},\ \Eprint {http://arxiv.org/abs/0710.4199} {arXiv:0710.4199 [hep-ex]} \BibitemShut {NoStop}%
\bibitem [{\citenamefont {Abazov}\ \emph {et~al.}(2008)\citenamefont {Abazov} \emph {et~al.}}]{D0:2008bs}%
  \BibitemOpen
  \bibfield  {author} {\bibinfo {author} {\bibfnamefont {V.~M.}\ \bibnamefont {Abazov}} \emph {et~al.} (\bibinfo {collaboration} {D0}),\ }\href {\doibase 10.1103/PhysRevLett.100.082002} {\bibfield  {journal} {\bibinfo  {journal} {Phys. Rev. Lett.}\ }\textbf {\bibinfo {volume} {100}},\ \bibinfo {pages} {082002} (\bibinfo {year} {2008})},\ \Eprint {http://arxiv.org/abs/0711.0319} {arXiv:0711.0319 [hep-ex]} \BibitemShut {NoStop}%
\bibitem [{\citenamefont {Aaij}\ \emph {et~al.}(2021{\natexlab{b}})\citenamefont {Aaij} \emph {et~al.}}]{LHCb:2020bsJ}%
  \BibitemOpen
  \bibfield  {author} {\bibinfo {author} {\bibfnamefont {R.}~\bibnamefont {Aaij}} \emph {et~al.} (\bibinfo {collaboration} {LHCb}),\ }\href {\doibase 10.1140/epjc/s10052-021-09305-3} {\bibfield  {journal} {\bibinfo  {journal} {Eur. Phys. J. C}\ }\textbf {\bibinfo {volume} {81}},\ \bibinfo {pages} {601} (\bibinfo {year} {2021}{\natexlab{b}})},\ \Eprint {http://arxiv.org/abs/2010.15931} {arXiv:2010.15931 [hep-ex]} \BibitemShut {NoStop}%
\bibitem [{\citenamefont {Aad}\ \emph {et~al.}(2014)\citenamefont {Aad} \emph {et~al.}}]{ATLAS:2014bc2S}%
  \BibitemOpen
  \bibfield  {author} {\bibinfo {author} {\bibfnamefont {G.}~\bibnamefont {Aad}} \emph {et~al.} (\bibinfo {collaboration} {ATLAS}),\ }\href {\doibase 10.1103/PhysRevLett.113.212004} {\bibfield  {journal} {\bibinfo  {journal} {Phys. Rev. Lett.}\ }\textbf {\bibinfo {volume} {113}},\ \bibinfo {pages} {212004} (\bibinfo {year} {2014})},\ \Eprint {http://arxiv.org/abs/1407.1032} {arXiv:1407.1032 [hep-ex]} \BibitemShut {NoStop}%
\bibitem [{\citenamefont {Sirunyan}\ \emph {et~al.}(2019)\citenamefont {Sirunyan} \emph {et~al.}}]{CMS:2019bc2S}%
  \BibitemOpen
  \bibfield  {author} {\bibinfo {author} {\bibfnamefont {A.~M.}\ \bibnamefont {Sirunyan}} \emph {et~al.} (\bibinfo {collaboration} {CMS}),\ }\href {\doibase 10.1103/PhysRevLett.122.132001} {\bibfield  {journal} {\bibinfo  {journal} {Phys. Rev. Lett.}\ }\textbf {\bibinfo {volume} {122}},\ \bibinfo {pages} {132001} (\bibinfo {year} {2019})},\ \Eprint {http://arxiv.org/abs/1902.00571} {arXiv:1902.00571 [hep-ex]} \BibitemShut {NoStop}%
\bibitem [{\citenamefont {Aad}\ \emph {et~al.}(2026)\citenamefont {Aad} \emph {et~al.}}]{ATLAS:2026abc}%
  \BibitemOpen
  \bibfield  {author} {\bibinfo {author} {\bibfnamefont {G.}~\bibnamefont {Aad}} \emph {et~al.} (\bibinfo {collaboration} {ATLAS}),\ }\href@noop {} {\bibfield  {journal} {\bibinfo  {journal} {Phys. Rev. Lett.}\ } (\bibinfo {year} {2026})},\ \bibinfo {note} {submitted},\ \Eprint {http://arxiv.org/abs/2605.16228} {arXiv:2605.16228 [hep-ex]} \BibitemShut {NoStop}%
\bibitem [{\citenamefont {Aaij}\ \emph {et~al.}(2025)\citenamefont {Aaij} \emph {et~al.}}]{LHCb:2025bcP}%
  \BibitemOpen
  \bibfield  {author} {\bibinfo {author} {\bibfnamefont {R.}~\bibnamefont {Aaij}} \emph {et~al.} (\bibinfo {collaboration} {LHCb}),\ }\href {\doibase 10.1103/PhysRevLett.135.231902} {\bibfield  {journal} {\bibinfo  {journal} {Phys. Rev. Lett.}\ }\textbf {\bibinfo {volume} {135}},\ \bibinfo {pages} {231902} (\bibinfo {year} {2025})}\BibitemShut {NoStop}%
\bibitem [{\citenamefont {Godfrey}\ and\ \citenamefont {Isgur}(1985)}]{Godfrey:1985xj}%
  \BibitemOpen
  \bibfield  {author} {\bibinfo {author} {\bibfnamefont {S.}~\bibnamefont {Godfrey}}\ and\ \bibinfo {author} {\bibfnamefont {N.}~\bibnamefont {Isgur}},\ }\href {\doibase 10.1103/PhysRevD.32.189} {\bibfield  {journal} {\bibinfo  {journal} {Phys. Rev. D}\ }\textbf {\bibinfo {volume} {32}},\ \bibinfo {pages} {189} (\bibinfo {year} {1985})}\BibitemShut {NoStop}%
\bibitem [{\citenamefont {Di~Pierro}\ and\ \citenamefont {Eichten}(2001)}]{DiPierro:2001hl}%
  \BibitemOpen
  \bibfield  {author} {\bibinfo {author} {\bibfnamefont {M.}~\bibnamefont {Di~Pierro}}\ and\ \bibinfo {author} {\bibfnamefont {E.}~\bibnamefont {Eichten}},\ }\href {\doibase 10.1103/PhysRevD.64.114004} {\bibfield  {journal} {\bibinfo  {journal} {Phys. Rev. D}\ }\textbf {\bibinfo {volume} {64}},\ \bibinfo {pages} {114004} (\bibinfo {year} {2001})},\ \Eprint {http://arxiv.org/abs/hep-ph/0104208} {arXiv:hep-ph/0104208} \BibitemShut {NoStop}%
\bibitem [{\citenamefont {Godfrey}(2004)}]{Godfrey:2004bc}%
  \BibitemOpen
  \bibfield  {author} {\bibinfo {author} {\bibfnamefont {S.}~\bibnamefont {Godfrey}},\ }\href {\doibase 10.1103/PhysRevD.70.054017} {\bibfield  {journal} {\bibinfo  {journal} {Phys. Rev. D}\ }\textbf {\bibinfo {volume} {70}},\ \bibinfo {pages} {054017} (\bibinfo {year} {2004})},\ \Eprint {http://arxiv.org/abs/hep-ph/0406228} {arXiv:hep-ph/0406228} \BibitemShut {NoStop}%
\bibitem [{\citenamefont {Godfrey}\ and\ \citenamefont {Moats}(2016)}]{Godfrey:2016charm}%
  \BibitemOpen
  \bibfield  {author} {\bibinfo {author} {\bibfnamefont {S.}~\bibnamefont {Godfrey}}\ and\ \bibinfo {author} {\bibfnamefont {K.}~\bibnamefont {Moats}},\ }\href {\doibase 10.1103/PhysRevD.93.034035} {\bibfield  {journal} {\bibinfo  {journal} {Phys. Rev. D}\ }\textbf {\bibinfo {volume} {93}},\ \bibinfo {pages} {034035} (\bibinfo {year} {2016})},\ \Eprint {http://arxiv.org/abs/1510.08305} {arXiv:1510.08305 [hep-ph]} \BibitemShut {NoStop}%
\bibitem [{\citenamefont {Godfrey}\ \emph {et~al.}(2016)\citenamefont {Godfrey}, \citenamefont {Moats},\ and\ \citenamefont {Swanson}}]{Godfrey:2016bs}%
  \BibitemOpen
  \bibfield  {author} {\bibinfo {author} {\bibfnamefont {S.}~\bibnamefont {Godfrey}}, \bibinfo {author} {\bibfnamefont {K.}~\bibnamefont {Moats}}, \ and\ \bibinfo {author} {\bibfnamefont {E.~S.}\ \bibnamefont {Swanson}},\ }\href {\doibase 10.1103/PhysRevD.94.054025} {\bibfield  {journal} {\bibinfo  {journal} {Phys. Rev. D}\ }\textbf {\bibinfo {volume} {94}},\ \bibinfo {pages} {054025} (\bibinfo {year} {2016})},\ \Eprint {http://arxiv.org/abs/1607.02169} {arXiv:1607.02169 [hep-ph]} \BibitemShut {NoStop}%
\bibitem [{\citenamefont {Ebert}\ \emph {et~al.}(2003)\citenamefont {Ebert}, \citenamefont {Faustov},\ and\ \citenamefont {Galkin}}]{Ebert:2002pp}%
  \BibitemOpen
  \bibfield  {author} {\bibinfo {author} {\bibfnamefont {D.}~\bibnamefont {Ebert}}, \bibinfo {author} {\bibfnamefont {R.~N.}\ \bibnamefont {Faustov}}, \ and\ \bibinfo {author} {\bibfnamefont {V.~O.}\ \bibnamefont {Galkin}},\ }\href {\doibase 10.1103/PhysRevD.67.014027} {\bibfield  {journal} {\bibinfo  {journal} {Phys. Rev. D}\ }\textbf {\bibinfo {volume} {67}},\ \bibinfo {pages} {014027} (\bibinfo {year} {2003})},\ \Eprint {http://arxiv.org/abs/hep-ph/0210381} {arXiv:hep-ph/0210381} \BibitemShut {NoStop}%
\bibitem [{\citenamefont {Ebert}\ \emph {et~al.}(2010)\citenamefont {Ebert}, \citenamefont {Faustov},\ and\ \citenamefont {Galkin}}]{Ebert:2009ua}%
  \BibitemOpen
  \bibfield  {author} {\bibinfo {author} {\bibfnamefont {D.}~\bibnamefont {Ebert}}, \bibinfo {author} {\bibfnamefont {R.~N.}\ \bibnamefont {Faustov}}, \ and\ \bibinfo {author} {\bibfnamefont {V.~O.}\ \bibnamefont {Galkin}},\ }\href {\doibase 10.1140/epjc/s10052-010-1233-6} {\bibfield  {journal} {\bibinfo  {journal} {Eur. Phys. J. C}\ }\textbf {\bibinfo {volume} {66}},\ \bibinfo {pages} {197} (\bibinfo {year} {2010})},\ \Eprint {http://arxiv.org/abs/0910.5612} {arXiv:0910.5612 [hep-ph]} \BibitemShut {NoStop}%
\bibitem [{\citenamefont {Ni}\ \emph {et~al.}(2022)\citenamefont {Ni}, \citenamefont {Li},\ and\ \citenamefont {Zhong}}]{Ni:2022cs}%
  \BibitemOpen
  \bibfield  {author} {\bibinfo {author} {\bibfnamefont {R.-H.}\ \bibnamefont {Ni}}, \bibinfo {author} {\bibfnamefont {Q.}~\bibnamefont {Li}}, \ and\ \bibinfo {author} {\bibfnamefont {X.-H.}\ \bibnamefont {Zhong}},\ }\href {\doibase 10.1103/PhysRevD.105.056006} {\bibfield  {journal} {\bibinfo  {journal} {Phys. Rev. D}\ }\textbf {\bibinfo {volume} {105}},\ \bibinfo {pages} {056006} (\bibinfo {year} {2022})},\ \Eprint {http://arxiv.org/abs/2110.05024} {arXiv:2110.05024 [hep-ph]} \BibitemShut {NoStop}%
\bibitem [{\citenamefont {Luo}\ \emph {et~al.}(2021)\citenamefont {Luo}, \citenamefont {Chen}, \citenamefont {Liu},\ and\ \citenamefont {Matsuki}}]{Luo:2021dvj}%
  \BibitemOpen
  \bibfield  {author} {\bibinfo {author} {\bibfnamefont {S.-Q.}\ \bibnamefont {Luo}}, \bibinfo {author} {\bibfnamefont {B.}~\bibnamefont {Chen}}, \bibinfo {author} {\bibfnamefont {X.}~\bibnamefont {Liu}}, \ and\ \bibinfo {author} {\bibfnamefont {T.}~\bibnamefont {Matsuki}},\ }\href {\doibase 10.1103/PhysRevD.103.074027} {\bibfield  {journal} {\bibinfo  {journal} {Phys. Rev. D}\ }\textbf {\bibinfo {volume} {103}},\ \bibinfo {pages} {074027} (\bibinfo {year} {2021})},\ \Eprint {http://arxiv.org/abs/2102.00679} {arXiv:2102.00679 [hep-ph]} \BibitemShut {NoStop}%
\bibitem [{\citenamefont {Zhang}\ \emph {et~al.}(2024)\citenamefont {Zhang}, \citenamefont {Liu}, \citenamefont {Luo}, \citenamefont {Chen},\ and\ \citenamefont {Guo}}]{Zhang:2024usz}%
  \BibitemOpen
  \bibfield  {author} {\bibinfo {author} {\bibfnamefont {Z.-L.}\ \bibnamefont {Zhang}}, \bibinfo {author} {\bibfnamefont {Z.-W.}\ \bibnamefont {Liu}}, \bibinfo {author} {\bibfnamefont {S.-Q.}\ \bibnamefont {Luo}}, \bibinfo {author} {\bibfnamefont {P.}~\bibnamefont {Chen}}, \ and\ \bibinfo {author} {\bibfnamefont {Z.-H.}\ \bibnamefont {Guo}},\ }\href {\doibase 10.1103/PhysRevD.110.094037} {\bibfield  {journal} {\bibinfo  {journal} {Phys. Rev. D}\ }\textbf {\bibinfo {volume} {110}},\ \bibinfo {pages} {094037} (\bibinfo {year} {2024})},\ \Eprint {http://arxiv.org/abs/2409.05337} {arXiv:2409.05337 [hep-ph]} \BibitemShut {NoStop}%
\bibitem [{\citenamefont {Gonzalez}\ \emph {et~al.}(2003)\citenamefont {Gonzalez}, \citenamefont {Valcarce}, \citenamefont {Garcilazo},\ and\ \citenamefont {Vijande}}]{Gonzalez:2003scr}%
  \BibitemOpen
  \bibfield  {author} {\bibinfo {author} {\bibfnamefont {P.}~\bibnamefont {Gonzalez}}, \bibinfo {author} {\bibfnamefont {A.}~\bibnamefont {Valcarce}}, \bibinfo {author} {\bibfnamefont {H.}~\bibnamefont {Garcilazo}}, \ and\ \bibinfo {author} {\bibfnamefont {J.}~\bibnamefont {Vijande}},\ }\href {\doibase 10.1103/PhysRevD.68.034007} {\bibfield  {journal} {\bibinfo  {journal} {Phys. Rev. D}\ }\textbf {\bibinfo {volume} {68}},\ \bibinfo {pages} {034007} (\bibinfo {year} {2003})},\ \Eprint {http://arxiv.org/abs/hep-ph/0307310} {arXiv:hep-ph/0307310} \BibitemShut {NoStop}%
\bibitem [{\citenamefont {Song}\ \emph {et~al.}(2015)\citenamefont {Song}, \citenamefont {Chen}, \citenamefont {Liu},\ and\ \citenamefont {Matsuki}}]{Song:2015nia}%
  \BibitemOpen
  \bibfield  {author} {\bibinfo {author} {\bibfnamefont {Q.-T.}\ \bibnamefont {Song}}, \bibinfo {author} {\bibfnamefont {D.-Y.}\ \bibnamefont {Chen}}, \bibinfo {author} {\bibfnamefont {X.}~\bibnamefont {Liu}}, \ and\ \bibinfo {author} {\bibfnamefont {T.}~\bibnamefont {Matsuki}},\ }\href {\doibase 10.1103/PhysRevD.91.054031} {\bibfield  {journal} {\bibinfo  {journal} {Phys. Rev. D}\ }\textbf {\bibinfo {volume} {91}},\ \bibinfo {pages} {054031} (\bibinfo {year} {2015})},\ \Eprint {http://arxiv.org/abs/1501.03575} {arXiv:1501.03575 [hep-ph]} \BibitemShut {NoStop}%
\bibitem [{\citenamefont {Wang}\ \emph {et~al.}(2018)\citenamefont {Wang}, \citenamefont {Sun}, \citenamefont {Liu},\ and\ \citenamefont {Matsuki}}]{Wang:2018rjg}%
  \BibitemOpen
  \bibfield  {author} {\bibinfo {author} {\bibfnamefont {J.-Z.}\ \bibnamefont {Wang}}, \bibinfo {author} {\bibfnamefont {Z.-F.}\ \bibnamefont {Sun}}, \bibinfo {author} {\bibfnamefont {X.}~\bibnamefont {Liu}}, \ and\ \bibinfo {author} {\bibfnamefont {T.}~\bibnamefont {Matsuki}},\ }\href {\doibase 10.1140/epjc/s10052-018-6372-1} {\bibfield  {journal} {\bibinfo  {journal} {Eur. Phys. J. C}\ }\textbf {\bibinfo {volume} {78}},\ \bibinfo {pages} {915} (\bibinfo {year} {2018})},\ \Eprint {http://arxiv.org/abs/1802.04938} {arXiv:1802.04938 [hep-ph]} \BibitemShut {NoStop}%
\bibitem [{\citenamefont {Wang}\ \emph {et~al.}(2019)\citenamefont {Wang}, \citenamefont {Chen}, \citenamefont {Liu},\ and\ \citenamefont {Matsuki}}]{Wang:2019mhs}%
  \BibitemOpen
  \bibfield  {author} {\bibinfo {author} {\bibfnamefont {J.-Z.}\ \bibnamefont {Wang}}, \bibinfo {author} {\bibfnamefont {D.-Y.}\ \bibnamefont {Chen}}, \bibinfo {author} {\bibfnamefont {X.}~\bibnamefont {Liu}}, \ and\ \bibinfo {author} {\bibfnamefont {T.}~\bibnamefont {Matsuki}},\ }\href {\doibase 10.1103/PhysRevD.99.114003} {\bibfield  {journal} {\bibinfo  {journal} {Phys. Rev. D}\ }\textbf {\bibinfo {volume} {99}},\ \bibinfo {pages} {114003} (\bibinfo {year} {2019})},\ \Eprint {http://arxiv.org/abs/1903.07115} {arXiv:1903.07115 [hep-ph]} \BibitemShut {NoStop}%
\bibitem [{\citenamefont {Wang}\ \emph {et~al.}(2020)\citenamefont {Wang}, \citenamefont {Qian}, \citenamefont {Liu},\ and\ \citenamefont {Matsuki}}]{Wang:2020prx}%
  \BibitemOpen
  \bibfield  {author} {\bibinfo {author} {\bibfnamefont {J.-Z.}\ \bibnamefont {Wang}}, \bibinfo {author} {\bibfnamefont {R.-Q.}\ \bibnamefont {Qian}}, \bibinfo {author} {\bibfnamefont {X.}~\bibnamefont {Liu}}, \ and\ \bibinfo {author} {\bibfnamefont {T.}~\bibnamefont {Matsuki}},\ }\href {\doibase 10.1103/PhysRevD.101.034001} {\bibfield  {journal} {\bibinfo  {journal} {Phys. Rev. D}\ }\textbf {\bibinfo {volume} {101}},\ \bibinfo {pages} {034001} (\bibinfo {year} {2020})},\ \Eprint {http://arxiv.org/abs/2001.00175} {arXiv:2001.00175 [hep-ph]} \BibitemShut {NoStop}%
\bibitem [{\citenamefont {Bai}\ \emph {et~al.}(2027)\citenamefont {Bai}, \citenamefont {Chen}, \citenamefont {Huang}, \citenamefont {Liu}, \citenamefont {Luo},\ and\ \citenamefont {Wang}}]{Bai:2026atm}%
  \BibitemOpen
  \bibfield  {author} {\bibinfo {author} {\bibfnamefont {Z.-Y.}\ \bibnamefont {Bai}}, \bibinfo {author} {\bibfnamefont {D.-Y.}\ \bibnamefont {Chen}}, \bibinfo {author} {\bibfnamefont {Q.}~\bibnamefont {Huang}}, \bibinfo {author} {\bibfnamefont {X.}~\bibnamefont {Liu}}, \bibinfo {author} {\bibfnamefont {S.-Q.}\ \bibnamefont {Luo}}, \ and\ \bibinfo {author} {\bibfnamefont {J.-Z.}\ \bibnamefont {Wang}},\ }\href {\doibase 10.1016/j.physrep.2026.08.002} {\bibfield  {journal} {\bibinfo  {journal} {Phys. Rept.}\ }\textbf {\bibinfo {volume} {1204}},\ \bibinfo {pages} {1} (\bibinfo {year} {2027})},\ \Eprint {http://arxiv.org/abs/2602.19887} {arXiv:2602.19887 [hep-ph]} \BibitemShut {NoStop}%
\bibitem [{\citenamefont {Ding}\ \emph {et~al.}(1993)\citenamefont {Ding}, \citenamefont {Chao},\ and\ \citenamefont {Qin}}]{Ding:1993uy}%
  \BibitemOpen
  \bibfield  {author} {\bibinfo {author} {\bibfnamefont {Y.-B.}\ \bibnamefont {Ding}}, \bibinfo {author} {\bibfnamefont {K.-T.}\ \bibnamefont {Chao}}, \ and\ \bibinfo {author} {\bibfnamefont {D.-H.}\ \bibnamefont {Qin}},\ }\href {\doibase 10.1088/0256-307X/10/8/004} {\bibfield  {journal} {\bibinfo  {journal} {Chin. Phys. Lett.}\ }\textbf {\bibinfo {volume} {10}},\ \bibinfo {pages} {460} (\bibinfo {year} {1993})}\BibitemShut {NoStop}%
\bibitem [{\citenamefont {Li}\ and\ \citenamefont {Chao}(2009)}]{Li:2009nr}%
  \BibitemOpen
  \bibfield  {author} {\bibinfo {author} {\bibfnamefont {B.-Q.}\ \bibnamefont {Li}}\ and\ \bibinfo {author} {\bibfnamefont {K.-T.}\ \bibnamefont {Chao}},\ }\href {\doibase 10.1088/0253-6102/52/4/20} {\bibfield  {journal} {\bibinfo  {journal} {Commun. Theor. Phys.}\ }\textbf {\bibinfo {volume} {52}},\ \bibinfo {pages} {653} (\bibinfo {year} {2009})},\ \Eprint {http://arxiv.org/abs/0909.1369} {arXiv:0909.1369 [hep-ph]} \BibitemShut {NoStop}%
\bibitem [{\citenamefont {Deng}\ \emph {et~al.}(2024)\citenamefont {Deng}, \citenamefont {Ni}, \citenamefont {Li},\ and\ \citenamefont {Zhong}}]{Deng:2023mza}%
  \BibitemOpen
  \bibfield  {author} {\bibinfo {author} {\bibfnamefont {Q.}~\bibnamefont {Deng}}, \bibinfo {author} {\bibfnamefont {R.-H.}\ \bibnamefont {Ni}}, \bibinfo {author} {\bibfnamefont {Q.}~\bibnamefont {Li}}, \ and\ \bibinfo {author} {\bibfnamefont {X.-H.}\ \bibnamefont {Zhong}},\ }\href {\doibase 10.1103/PhysRevD.110.056034} {\bibfield  {journal} {\bibinfo  {journal} {Phys. Rev. D}\ }\textbf {\bibinfo {volume} {110}},\ \bibinfo {pages} {056034} (\bibinfo {year} {2024})},\ \Eprint {http://arxiv.org/abs/2312.10296} {arXiv:2312.10296 [hep-ph]} \BibitemShut {NoStop}%
\bibitem [{\citenamefont {Chen}\ and\ \citenamefont {Tan}(2024)}]{Chen:2024ukv}%
  \BibitemOpen
  \bibfield  {author} {\bibinfo {author} {\bibfnamefont {X.}~\bibnamefont {Chen}}\ and\ \bibinfo {author} {\bibfnamefont {Y.}~\bibnamefont {Tan}},\ }\href {\doibase 10.1088/1674-1137/ad53bd} {\bibfield  {journal} {\bibinfo  {journal} {Chin. Phys. C}\ }\textbf {\bibinfo {volume} {48}},\ \bibinfo {pages} {093107} (\bibinfo {year} {2024})},\ \Eprint {http://arxiv.org/abs/2406.00957} {arXiv:2406.00957 [hep-ph]} \BibitemShut {NoStop}%
\bibitem [{\citenamefont {Zhang}\ \emph {et~al.}(2026)\citenamefont {Zhang}, \citenamefont {Luo},\ and\ \citenamefont {Liu}}]{Zhang:2026yif}%
  \BibitemOpen
  \bibfield  {author} {\bibinfo {author} {\bibfnamefont {W.-X.}\ \bibnamefont {Zhang}}, \bibinfo {author} {\bibfnamefont {S.-Q.}\ \bibnamefont {Luo}}, \ and\ \bibinfo {author} {\bibfnamefont {X.}~\bibnamefont {Liu}},\ }\href@noop {} {\  (\bibinfo {year} {2026})},\ \Eprint {http://arxiv.org/abs/2608.16875} {arXiv:2608.16875 [hep-ph]} \BibitemShut {NoStop}%
\bibitem [{\citenamefont {Capstick}\ and\ \citenamefont {Isgur}(1986)}]{Capstick:1986ter}%
  \BibitemOpen
  \bibfield  {author} {\bibinfo {author} {\bibfnamefont {S.}~\bibnamefont {Capstick}}\ and\ \bibinfo {author} {\bibfnamefont {N.}~\bibnamefont {Isgur}},\ }\href {\doibase 10.1103/physrevd.34.2809} {\bibfield  {journal} {\bibinfo  {journal} {Phys. Rev. D}\ }\textbf {\bibinfo {volume} {34}},\ \bibinfo {pages} {2809} (\bibinfo {year} {1986})}\BibitemShut {NoStop}%
\bibitem [{\citenamefont {Hiyama}\ \emph {et~al.}(2003)\citenamefont {Hiyama}, \citenamefont {Kino},\ and\ \citenamefont {Kamimura}}]{Hiyama:2003cu}%
  \BibitemOpen
  \bibfield  {author} {\bibinfo {author} {\bibfnamefont {E.}~\bibnamefont {Hiyama}}, \bibinfo {author} {\bibfnamefont {Y.}~\bibnamefont {Kino}}, \ and\ \bibinfo {author} {\bibfnamefont {M.}~\bibnamefont {Kamimura}},\ }\href {\doibase 10.1016/S0146-6410(03)90015-9} {\bibfield  {journal} {\bibinfo  {journal} {Prog. Part. Nucl. Phys.}\ }\textbf {\bibinfo {volume} {51}},\ \bibinfo {pages} {223} (\bibinfo {year} {2003})}\BibitemShut {NoStop}%
\bibitem [{\citenamefont {Hiyama}(2012)}]{Hiyama:2012sma}%
  \BibitemOpen
  \bibfield  {author} {\bibinfo {author} {\bibfnamefont {E.}~\bibnamefont {Hiyama}},\ }\href {\doibase 10.1093/ptep/pts015} {\bibfield  {journal} {\bibinfo  {journal} {PTEP}\ }\textbf {\bibinfo {volume} {2012}},\ \bibinfo {pages} {01A204} (\bibinfo {year} {2012})}\BibitemShut {NoStop}%
\bibitem [{\citenamefont {Luo}\ \emph {et~al.}(2020)\citenamefont {Luo}, \citenamefont {Chen}, \citenamefont {Liu},\ and\ \citenamefont {Liu}}]{Luo:2019qkm}%
  \BibitemOpen
  \bibfield  {author} {\bibinfo {author} {\bibfnamefont {S.-Q.}\ \bibnamefont {Luo}}, \bibinfo {author} {\bibfnamefont {B.}~\bibnamefont {Chen}}, \bibinfo {author} {\bibfnamefont {Z.-W.}\ \bibnamefont {Liu}}, \ and\ \bibinfo {author} {\bibfnamefont {X.}~\bibnamefont {Liu}},\ }\href {\doibase 10.1140/epjc/s10052-020-7874-1} {\bibfield  {journal} {\bibinfo  {journal} {Eur. Phys. J. C}\ }\textbf {\bibinfo {volume} {80}},\ \bibinfo {pages} {301} (\bibinfo {year} {2020})},\ \Eprint {http://arxiv.org/abs/1910.14545} {arXiv:1910.14545 [hep-ph]} \BibitemShut {NoStop}%
\bibitem [{\citenamefont {Duan}\ and\ \citenamefont {Liu}(2021)}]{Duan:2021pw}%
  \BibitemOpen
  \bibfield  {author} {\bibinfo {author} {\bibfnamefont {M.-X.}\ \bibnamefont {Duan}}\ and\ \bibinfo {author} {\bibfnamefont {X.}~\bibnamefont {Liu}},\ }\href {\doibase 10.1103/PhysRevD.104.074010} {\bibfield  {journal} {\bibinfo  {journal} {Phys. Rev. D}\ }\textbf {\bibinfo {volume} {104}},\ \bibinfo {pages} {074010} (\bibinfo {year} {2021})},\ \Eprint {http://arxiv.org/abs/2107.14438} {arXiv:2107.14438 [hep-ph]} \BibitemShut {NoStop}%
\bibitem [{\citenamefont {Man}\ \emph {et~al.}(2024)\citenamefont {Man}, \citenamefont {Shu}, \citenamefont {Liu},\ and\ \citenamefont {Chen}}]{Man:2024mvl}%
  \BibitemOpen
  \bibfield  {author} {\bibinfo {author} {\bibfnamefont {Z.-L.}\ \bibnamefont {Man}}, \bibinfo {author} {\bibfnamefont {C.-R.}\ \bibnamefont {Shu}}, \bibinfo {author} {\bibfnamefont {Y.-R.}\ \bibnamefont {Liu}}, \ and\ \bibinfo {author} {\bibfnamefont {H.}~\bibnamefont {Chen}},\ }\href {\doibase 10.1140/epjc/s10052-024-13132-7} {\bibfield  {journal} {\bibinfo  {journal} {Eur. Phys. J. C}\ }\textbf {\bibinfo {volume} {84}},\ \bibinfo {pages} {810} (\bibinfo {year} {2024})},\ \Eprint {http://arxiv.org/abs/2402.02765} {arXiv:2402.02765 [hep-ph]} \BibitemShut {NoStop}%
\bibitem [{\citenamefont {Man}\ \emph {et~al.}(2025{\natexlab{a}})\citenamefont {Man}, \citenamefont {Luo}, \citenamefont {Bai},\ and\ \citenamefont {Liu}}]{Man:2025zfu}%
  \BibitemOpen
  \bibfield  {author} {\bibinfo {author} {\bibfnamefont {Z.-L.}\ \bibnamefont {Man}}, \bibinfo {author} {\bibfnamefont {S.-Q.}\ \bibnamefont {Luo}}, \bibinfo {author} {\bibfnamefont {Z.-Y.}\ \bibnamefont {Bai}}, \ and\ \bibinfo {author} {\bibfnamefont {X.}~\bibnamefont {Liu}},\ }\href {\doibase 10.1016/j.physletb.2025.139644} {\bibfield  {journal} {\bibinfo  {journal} {Phys. Lett. B}\ }\textbf {\bibinfo {volume} {868}},\ \bibinfo {pages} {139644} (\bibinfo {year} {2025}{\natexlab{a}})},\ \Eprint {http://arxiv.org/abs/2502.08072} {arXiv:2502.08072 [hep-ph]} \BibitemShut {NoStop}%
\bibitem [{\citenamefont {Man}\ \emph {et~al.}(2025{\natexlab{b}})\citenamefont {Man}, \citenamefont {Luo},\ and\ \citenamefont {Liu}}]{Man:2025vmm}%
  \BibitemOpen
  \bibfield  {author} {\bibinfo {author} {\bibfnamefont {Z.-L.}\ \bibnamefont {Man}}, \bibinfo {author} {\bibfnamefont {S.-Q.}\ \bibnamefont {Luo}}, \ and\ \bibinfo {author} {\bibfnamefont {X.}~\bibnamefont {Liu}},\ }\href {\doibase 10.1103/76fd-njsy} {\bibfield  {journal} {\bibinfo  {journal} {Phys. Rev. D}\ }\textbf {\bibinfo {volume} {112}},\ \bibinfo {pages} {074025} (\bibinfo {year} {2025}{\natexlab{b}})},\ \Eprint {http://arxiv.org/abs/2507.18536} {arXiv:2507.18536 [hep-ph]} \BibitemShut {NoStop}%
\bibitem [{\citenamefont {Qian}\ and\ \citenamefont {Liu}(2025)}]{Qian:2025cc}%
  \BibitemOpen
  \bibfield  {author} {\bibinfo {author} {\bibfnamefont {R.-Q.}\ \bibnamefont {Qian}}\ and\ \bibinfo {author} {\bibfnamefont {X.}~\bibnamefont {Liu}},\ }\href {\doibase 10.1103/c5vs-2l1k} {\bibfield  {journal} {\bibinfo  {journal} {Phys. Rev. D}\ }\textbf {\bibinfo {volume} {112}},\ \bibinfo {pages} {L091502} (\bibinfo {year} {2025})},\ \Eprint {http://arxiv.org/abs/2509.17679} {arXiv:2509.17679 [hep-ph]} \BibitemShut {NoStop}%
\bibitem [{\citenamefont {Gonz{\'a}lez}(2015)}]{Gonzalez:2015hqa}%
  \BibitemOpen
  \bibfield  {author} {\bibinfo {author} {\bibfnamefont {P.}~\bibnamefont {Gonz{\'a}lez}},\ }\href {\doibase 10.1103/PhysRevD.92.014017} {\bibfield  {journal} {\bibinfo  {journal} {Phys. Rev. D}\ }\textbf {\bibinfo {volume} {92}},\ \bibinfo {pages} {014017} (\bibinfo {year} {2015})},\ \Eprint {http://arxiv.org/abs/1507.02397} {arXiv:1507.02397 [hep-ph]} \BibitemShut {NoStop}%
\bibitem [{\citenamefont {Brun}\ and\ \citenamefont {Rademakers}(1997)}]{ROOT_NIMA_1997}%
  \BibitemOpen
  \bibfield  {author} {\bibinfo {author} {\bibfnamefont {R.}~\bibnamefont {Brun}}\ and\ \bibinfo {author} {\bibfnamefont {F.}~\bibnamefont {Rademakers}},\ }\href {\doibase 10.1016/S0168-9002(97)00048-X} {\bibfield  {journal} {\bibinfo  {journal} {Nucl. Instrum. Meth. A}\ }\textbf {\bibinfo {volume} {389}},\ \bibinfo {pages} {81} (\bibinfo {year} {1997})}\BibitemShut {NoStop}%
\bibitem [{\citenamefont {SpaceXAI}(2026)}]{xaigrok}%
  \BibitemOpen
  \bibfield  {author} {\bibinfo {author} {\bibnamefont {SpaceXAI}},\ }\href {https://docs.x.ai/developers/models/grok-4.6} {\enquote {\bibinfo {title} {Grok-4.6},}\ } (\bibinfo {year} {2026})\BibitemShut {NoStop}%
\bibitem [{\citenamefont {Zhang}(2026)}]{gemstore}%
  \BibitemOpen
  \bibfield  {author} {\bibinfo {author} {\bibfnamefont {W.-X.}\ \bibnamefont {Zhang}},\ }\href {https://github.com/SerialCore/gemstore} {\enquote {\bibinfo {title} {Gemstore: Hadron spectroscopy simulation tools},}\ } (\bibinfo {year} {2026})\BibitemShut {NoStop}%
\end{thebibliography}
\end{document}